\documentclass[twocolumn,journal]{IEEEtran}
\usepackage[T1]{fontenc}
\usepackage[latin9]{inputenc}
\usepackage{color}
\usepackage{float}
\usepackage{booktabs}
\usepackage{amsmath}
\usepackage{amssymb}
\usepackage{graphicx}
\usepackage[numbers,numbers,sort&compress]{natbib}

\makeatletter

\providecommand{\tabularnewline}{\\}

\let\oldforeign@language\foreign@language
\DeclareRobustCommand{\foreign@language}[1]{%
  \lowercase{\oldforeign@language{#1}}}

\usepackage[ruled,vlined,linesnumbered]{algorithm2e}

\usepackage{graphicx}

\SetKwInput{KwInput}{Input}
\SetKwInput{KwOutput}{Output}

\usepackage{algorithmicx}
\usepackage{algpseudocode}
\usepackage{amsmath,amsfonts,amssymb}

\usepackage{array,booktabs,arydshln}
\usepackage{xcolor}
\usepackage{arydshln}

\usepackage{float} 

\usepackage[caption=false,font=footnotesize]{subfig}

\usepackage{tikz}

\usepackage{tcolorbox}

\newenvironment{alProblem}[1][htb]
	{
	\let\c@algocf\c@alproblem
	\begin{algorithm}[#1]%
}{\end{algorithm}}

\@ifundefined{showcaptionsetup}{}{%
 \PassOptionsToPackage{caption=false}{subfig}}
\usepackage{subfig}
\makeatother

\begin{document}
\title{Admission Control and Resource Reservation\\
for Prioritized Slice Requests with Guaranteed SLA under\textcolor{blue}{{}
}Uncertainties}
\author{Quang-Trung~Luu, Sylvaine~Kerboeuf, and~Michel~Kieffer \thanks{Quang-Trung Luu is with Nokia Bell Labs, 91620 Nozay, France, and
also with Laboratoire des Signaux et Syst\`emes, Paris-Saclay University
- CNRS - CentraleSup\'elec, 91192 Gif-sur-Yvette, France (e-mail:
quangtrung.luu@centralesupelec.fr).}\thanks{S.~Kerboeuf is with NSSR Lab, Nokia Bell Labs, 91629 Nozay, France
(e-mail: sylvaine.kerboeuf@ nokia-bell-labs.com).}\thanks{M.~Kieffer is with Laboratoire des Signaux et Syst\`emes, Paris-Saclay
University - CNRS - CentraleSup\'elec, 91192 Gif-sur-Yvette, France
(e-mail: michel.kieffer@l2s.centralesupelec.fr).}}
\markboth{Preprint}{Q.-T.~Luu \MakeLowercase{\emph{et al.}}: Admission Control and
Resource Reservation for Prioritized Slice Requests with Uncertainties}
\maketitle
\begin{abstract}
Network slicing has emerged as a key concept in 5G systems, allowing
Mobile Network Operators (MNOs) to build isolated logical networks
(slices) on top of shared infrastructure networks managed by Infrastructure
Providers (InP). Network slicing requires the assignment of infrastructure
network resources to virtual network components at slice activation
time and the adjustment of resources for slices under operation. Performing
these operations \textit{just-in-time}, on a best-effort basis, comes
with no guarantee on the availability of enough infrastructure resources
to meet slice requirements.

This paper proposes a prioritized admission control mechanism for
concurrent slices based on an infrastructure resource reservation
approach. The reservation accounts for the dynamic nature of slice
requests while being robust to uncertainties in slice resource demands.
Adopting the perspective of an InP, reservation schemes are proposed
that maximize the number of slices for which infrastructure resources
can be granted while minimizing the costs charged to the MNOs. This
requires the solution of a max-min optimization problem with a non-linear
cost function and non-linear constraints induced by the robustness
to uncertainties of demands and the limitation of the impact of reservation
on background services. The cost and the constraints are linearized
and several reduced-complexity strategies are proposed to solve the
slice admission control and resource reservation problem. Simulations
show that the proportion of admitted slices of different priority
levels can be adjusted by a differentiated selection of the delay
between the reception and the processing instants of a slice resource
request.
\end{abstract}


\IEEEpeerreviewmaketitle{}

\section{Introduction}

\IEEEPARstart{I}{n} the fifth-generation communication systems \cite{5GAmericas2016,IETF2017},
Network slicing (NS) aims at replacing the traditional \emph{one-size-fits-all}
network architecture. NS may address diverging requirements imposed
by verticals \cite{Ordonez-Lucena2017} while reducing operational
costs \cite{Liang2014,Rost2017}, thanks to its ability to provide
higher network flexibility. NS exploits network virtualization to
elastically allocate and reallocate infrastructure resources tailored
to the time-varying needs of various applications \cite{GSMA2017,Li2018}.
With NS, multiple \textit{slices}, \emph{i.e.}, \textit{customized},
\textit{isolated}, and \textit{service-dedicated} end-to-end logical
networks, can be established and operated simultaneously on a shared
physical infrastructure network, provided by one or several Infrastructure
Providers (InPs) \cite{Su2019a}.

Several authors, see, \emph{e.g.}, \cite{Huin2017,Wang2017,Su2019,Barakabitze2020},
have recently considered the resource allocation problem raised by
network slicing. This problem involves efficiently assigning infrastructure
network resources to virtual network components at (or just before)
slice activation time and dynamically adjusting these resources for
slices under operation to maximize resource utilization and minimize
operating costs. With such \emph{just-in-time} slice management, it
is difficult to guarantee the availability of enough infrastructure
resources at the deployment time and during the lifetime of a slice.
Slice admission control mechanisms have therefore been proposed to
prioritize, accept, possibly delay, or even reject demands for slices
\cite{Bega2017,Salvat2018,Noroozi2019,Bega2020,Ebrahimi2020,Han2020}.

Network slicing with guaranteed satisfaction of some Service Level
Agreement (SLA) is facilitated by adopting an infrastructure resource\emph{
}\textit{reservation} approach, as specified by \cite{3GPP2020,3GPP:TS-28.531},
rather than the \textit{just-in-time} slice resource management approach
considered, \emph{e.g.}, in \cite{Sun2019,Luu:unc}. Resource reservation
aims at determining whether enough infrastructure resources are (or
will be) available to satisfy a slice resource request (feasibility
check). The actual resource allocation may be done later, once the
reservation is successful and the slice request has been granted.

Nevertheless, slice resource reservation raises several challenging
problems: Slice requests are submitted by Service Providers (SPs)
at different time instants, with various activation delays, life durations,
and user demands fluctuating with time. The variety of services supported
by slices induces very different Quality of Service (QoS) requirements
\cite{Li2018}. Moreover, various constraints may be imposed by the
different network segments on which slices have to be deployed \cite{Salvat2018,Barakabitze2020}.
For example, coverage constraints are imposed by slices involving
the radio access network \cite{Luu:cov:icc}. Additionally, several
sources of uncertainty have to be considered in a reservation approach,
\emph{e.g.}, the number of slice users, the hardly predictable user
locations \cite{Richart2016}, and the time-varying per-user resource
requirements. Consequently, enough infrastructure resources should
be reserved for each slice to guarantee an adequate QoS specified
in the SLA and provide robustness against uncertainties. Too many
infrastructure resources should not be reserved too, in order to reduce
costs and leave resources to concurrent slices.

\noindent \textit{Contributions}. This work considers a model where
SPs submit slice service requests (possibly largely before their activation)
to some Mobile Network Operator (MNO). The MNO and the InP are considered
as two separate entities, possibly belonging to the same company.
The MNO evaluates the amount of resources required to operate each
slice efficiently and submits slice resource reservation requests
to an InP. The InP has to determine whether it is able to book, as
much in advance as possible, enough infrastructure resources to ensure
that the MNO will have access to enough and properly located infrastructure
resources with service characteristics as stated in some SLA. The
first contribution of this paper is a slice admission control and
resource reservation framework able to provide a probabilistic guarantee
related to SLA satisfaction. Overbooking of resources by the InP is
allowed, as in \cite{Salvat2018}, but the probability of SLA non-satisfaction
is bounded, with a controlled bound. The proposed method to reserve
infrastructure resources for concurrent slices accounts for the dynamic
nature of slice requests (including their arrival, activation, and
deactivation times while being robust against the uncertainties in
the number of users and the amount of resource employed by a typical
user). We adopt the perspective of an InP and propose an approach
where the InP tries to find the resource reservation scheme which
maximizes the amount of slices for which the reservation is successful
while minimizing the resource operation costs charged to the MNOs.
The InP decides then to accept or reject each slice resource request.

The processing of a slice request submitted largely before its activation
can be anticipated by the MNO who may check its feasibility in advance
with one or several InPs. The second contribution of this paper is
a process that anticipates more or less this processing depending
on the priority level of the slice requests. Slices can be admitted,
possibly largely before their activation, when enough infrastructure
resources are reserved for meeting their QoS requirements. When this
condition is not satisfied, the slice resource request is not granted
and MNOs may address their slice resource request to alternative InPs.
The proposed approach is consistent with the 3GPP views of the management
aspects of network slicing \cite{3GPP2020,3GPP:TS-28.531}. The proposed
slice reservation and admission control takes place in the \emph{network
environment preparation} task of the \emph{preparation} phase. In
this phase, the design and capacity planning of network slices, the
on-boarding and evaluation of required network functions, and the
reservation of infrastructure resources have to be done before the
creation and activation of network slice instances, which belong to
the \emph{commissioning} and \emph{operation} phases (see Figure~\ref{dy:fig:Intro:SliceLifeCycle}).

The third contribution of this paper is to formulate the processing
of concurrent slice resource reservation requests and their admission
control as a max-min optimization problem. Its solution provides the
InP maximum earnings for granted slice requests and is also appropriate
for the MNO in terms of charged reservation costs. Reservation and
adaptation costs are introduced to charge variations in slice resource
demands. A nonlinear objective function is then obtained. The robustness
to uncertainties in the demands and the limitation of the impact of
reservation on background services is considered as in \cite{Luu:unc}.
This introduces nonlinear constraints in the max-min optimization
problem. After linearizing the cost and some of the constraints, several
reduced-complexity reservation strategies are proposed to solve the
problem of slice resource reservation with dynamic resource demands.
\begin{center}
\begin{figure}[h]
\begin{centering}
\includegraphics[width=1\columnwidth]{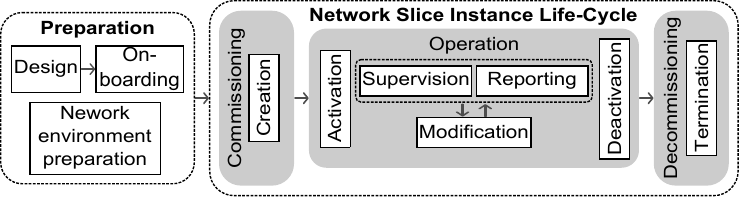}
\par\end{centering}
\caption{3GPP view on network slicing managements aspects \cite{3GPP2020}.\label{dy:fig:Intro:SliceLifeCycle}}
\end{figure}
\vspace{-0.5cm}
\par\end{center}

The remainder of this paper is organized as follows. Section~\ref{dy:sec:Related-work}
presents some related work. In Section~\ref{sec:Problem-Statement},
we describe the problem statement, in which the system model is detailed.
Section~\ref{dy:sec:Reservation-Approaches} introduces the proposed
approaches to efficiently reserve resources for concurrent slices,
while being robust to the dynamic nature of slice requests and to
the uncertainties related to infrastructure and slice parameters.
Numerical results are then provided in Section~\ref{dy:sec:Evaluation}
to evaluate the performance of the proposed reservation and admission
control approaches. Finally, Section~\ref{dy:sec:Conclusions} draws
some conclusions and perspectives.

\section{Related Work \label{dy:sec:Related-work}}

In network slicing, a slice is composed of one or multiple Service
Function Chains (SFCs) of different types. An SFC consists of an ordered
set of interconnected Virtual Network Functions\textit{\emph{ (}}VNFs)
describing the processing applied to data flows related to a given
service.

Many papers have addressed the problem of slice/SFC resource allocation
with uncertain or time-varying requirements and available physical
resources, see, \textit{e.g}., \cite{Huin2017,Wang2017,Coniglio2015,Mireslami2019,Baumgartner2018,Fendt2019}.
Conservative strategies allocating resources considering worst-case
peak traffic conditions \cite{Huin2017,Wang2017} are costly and usually
lead to an inefficient utilization of resources. In \cite{Baumgartner2018,Mireslami2019,Fendt2019},
an adjustable safety factor is considered to give a probabilistic
guarantee of resource availability, \textit{e.g}., ensuring that every
slice benefits from sufficient infrastructure resources with a certain
probability. Inspired by the approach in \cite{Fendt2019}, \cite{Luu:unc}
has extended the approach introduced in \cite{Luu:glo} to account
for the impact of resource reservation on background services.

The dynamic nature of slice/SFC requests is taken into account in
\cite{Liu2017,Sun2019,Wang2019,Huynh2019}. In \cite{Liu2017}, a
dynamic resource allocation for SFCs is investigated. The deployment
of newly arrived SFCs and readjustment of in-service SFCs are taken
into account. An Integer Linear Program (ILP) formulation is used
to address the dynamic deployment problem, aiming at minimizing the
cost of VNF deployment and migration. A pre-calculation of all possible
routing paths has to be performed in advance, which requires some
computational effort before using the deployment algorithm. In \cite{Sun2019},
the adaptive adjustment of allocated resources of each slice is enabled
after each decision time period (slicing time). An hybrid slice reconfiguration
framework is introduced in \cite{Wang2019}. The slice can be reconfigured
either within small time intervals for individual slices, or within
large time intervals to readjust resource allocation of multiple slices.
A deep-learning approach is adopted in \cite{Huynh2019} for dynamic
slice resource allocation, with the aim to maximize the long-term
revenue of the network provider. Uncertainties related to the slice
requests and occupation time are considered. Nevertheless, slices
are regarded as a whole, \textit{i.e}., not made up of multiple elements
(\textit{e.g}., VNFs), which somewhat over-simplifies the problem
of slice resource allocation.

Slice Admission control (SAC) mechanisms have been developed recently
\cite{Noroozi2019,Ebrahimi2020,Salvat2018,Han2020,Bega2017,Bega2020}
to address issues related to the unavailability of enough resources
to satisfy all slice requests. A yield-driven approach is proposed
in \cite{Salvat2018}, assuming that the MNO manages the infrastructure
resources and decides to accept or reject slice requests in order
to maximize the revenue obtained from the SPs. Resource overbooking
is allowed and a penalty in case of non-satisfaction is considered
in the optimization process. Nevertheless, there is no control on
the level of satisfaction of the SLA requirements imposed by the service
provider due to the overbooking possibility. Consequently, a slice
request may be granted with a penalty very close to the revenue provided
by the slice. Slice requests in \cite{Salvat2018} are assumed to
be processed as they arrive, without considering any explicit prioritization
between requests. Prioritized processing results only implicitly from
the differences in the revenue and penalty associated to the various
requests.

In \cite{Noroozi2019}, SAC is formulated as a boolean linear program
and a two-step sub-optimal algorithm based on variants of the knapsack
problem is proposed to alleviate the complexity. Admission is done
for slices with highest profit considering first the RAN and \emph{aggregated}
core network resources. In the second step, the core network resources
are considered without any aggregation to determine whether a slice
deployment is possible.

In \cite{Ebrahimi2020}, SAC and resource allocation are performed
jointly, to minimize the power consumption of cloud nodes and network
bandwidth of the infrastructure provider. Transmission delay is accounted
in the slice SLA. Some elastic variables are introduced in an ILP
formulation to extend the bounds on some constraints. They help determining
when resources may be lacking, in which case slices are rejected starting
from those with the highest requirements in terms of resource. Nevertheless,
the dynamics of slice requests (time of arrival, slice duration) and
the variation of slice resource demands during their life time are
not considered in \cite{Noroozi2019} and \cite{Ebrahimi2020}.

The dynamics of slice requests is considered by \cite{Han2020} in
the SAC problem. If not accepted, a request is queued for being potentially
served later. The case of impatient tenants, who may leave their queues
before being served, is taken into account. Nevertheless, neither
the dynamics of resource demands within each slice, nor the activation
time of a slice are accounted for. Moreover, infrastructure resources
of each type are fully aggregated. As opposed to \cite{Ebrahimi2020}
and to our work, none of the details about the structure of the slice
and of the infrastructure are taken into account in the resource model.
Consequently, the proposed mechanism does not allow to reserve nor
allocate resource to the slice in addition to admission control.

Online SAC is considered in \cite{Bega2017} and \cite{Bega2020}
leveraging on machine learning approaches. The aim is to maximize
the revenue of the InP while guaranteeing the SLAs of the admitted
slices. Both papers focus on radio resources of base stations. In
\cite{Bega2017}, two different types of slices are considered to
account for elastic and inelastic traffic. An admissibility region
is determined first, indicating the maximum number of slices that
the system can support without breaking the SLAs. Both works formalize
the admission control problem into a semi MDP and derived the optimal
policy obtained when the request arrival parameters are known. The
approach has a high computation cost and is off-line (requires system
parameters to be known \emph{a priori}). An alternative Q-learning
approach is proposed in \cite{Bega2017} to adapt to changing environments
while achieving close to optimal performance. In \cite{Bega2020},
a deep reinforcement learning method is developed to overcome the
scalability issue of the Q-learning approach.

These works consider tenants submitting slice requests for an immediate
deployment, contrary to our work, where slice requests are assumed
to be submitted for an immediate or future deployment, which permits
the development of a resource reservation strategy.

\section{Problem Statement and Notations\label{sec:Problem-Statement}}

A typical network slicing system involves several entities: one or
several InPs, MNOs, and SPs (also known as slice tenants), as depicted
in Figure~\ref{dy:fig:System-architecture} \cite{Liang2014}. InPs
own and manage the wireless and wired infrastructure such as the cell
sites, the fronthaul and backhaul networks, and data centers. Section~\ref{dy:subsec:Network-Model}
details the considered model for the infrastructure network. MNOs
lease resources from InPs to set up and manage slices. SPs then exploit
the slices supplied by MNOs and provide their customers with the required
services running within the slices. An SP forwards a slice service
booking request to an MNO within an SLA denoted SM-SLA in what follows.
\begin{center}
\begin{figure}[tbh]
\begin{centering}
\includegraphics[width=0.9\columnwidth]{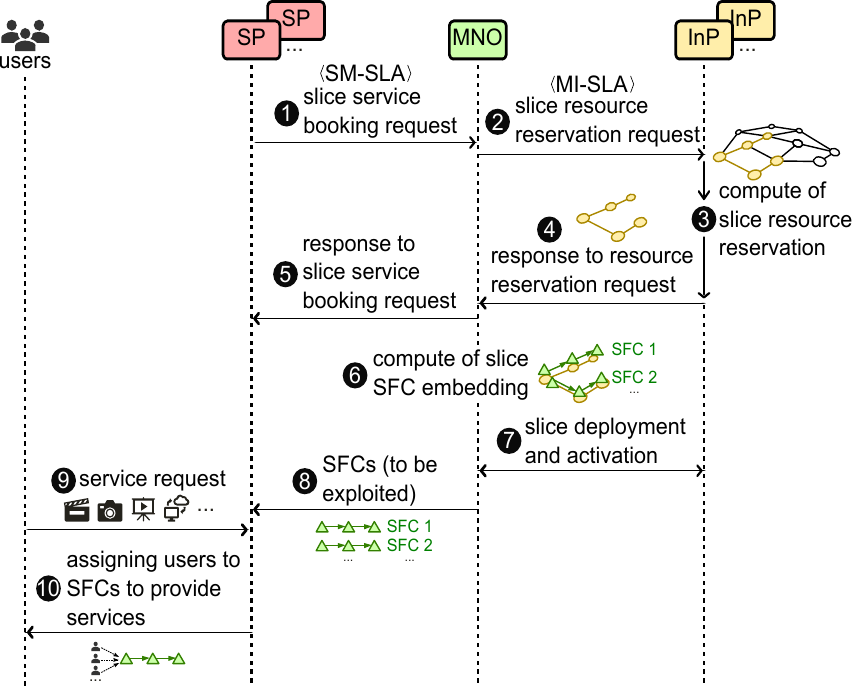}
\par\end{centering}
\caption{Network slicing entities and their SLA-based relationships.\label{dy:fig:System-architecture}}
\end{figure}
\vspace{-0.5cm}
\par\end{center}

The SM-SLA describes at a high level of abstraction characteristics
of the service with the desired QoS. These characteristics may be
time-varying due, \emph{e.g.}, to user mobility, see Section~\ref{dy:subsec:Request-Arrivals}.
In this paper, one considers SM-SLAs composed of: (\textit{i}) the
target number of users/devices to be supported by the slice, (\textit{ii})
a description of the characteristics of the service and of the way
typical user/device employs it, and (\textit{iii}) a target Service
Satisfaction Probability (SSP) $\underline{p}^{\text{sp}}$. Since
the target number of users is usually uncertain, it is described by
a random variable with a known probability mass function (pmf). The
MNO has to provide a slice able to serve user demands with a probability
of at least $\underline{p}^{\text{sp}}$. In addition, several time
intervals may be considered in the SM-SLA, intervals over each of
which the target number of users, the service characteristics, and
the probability of satisfaction are assumed invariant. They may vary
from a time interval to the next one. These time intervals translate,\textit{
e.g}., day and night variations of user demands and last between tens
of minutes to hours. It is the responsibility of the SP and MNO to
properly scale the requirements expressed in the SM-SLA by considering,
for example, similar services deployed in the past. Following the
3GPP approach (cf. Figure 4.8.1 of \cite{3GPP2020}), the MNO is in
charge of the slice admission control via assessing the feasibility
of the SP's request.

The MNO translates the SP high-level demands into SFCs able to fulfill
the service requirements. Based on the characteristics of the service
and of its usage, the MNO describes how a given user/device consumes
the slice (SFCs) resources. To characterize the variability over time
and among users of these demands, we assume that the MNO considers
a probabilistic description of the consumption of slice resources
by a typical user.

In what follows, one assumes that the MNO and the InP are two distinct
entities (possibly belonging to a single stakeholder but having their
domain responsibilities, like, \emph{e.g.}, two business divisions
of the same organization). In this case, the MNO submits resource
reservation requests to one or several InPs upon the arrival of slice
booking requests. Section~\ref{dy:subsec:Slice-Prov-Request} provides
a model of the requests for slice resource reservation sent by the
MNO to an InP and of their associated costs. Each resource reservation
request contains the description of the resource demand characteristics
(the SSP constraint is translated into deterministic requirements,
see Section~\ref{dy:subsec:Relax:Proba}) as well as the slice priority
class as part of an SLA between them (MI-SLA), see Section~\ref{dy:subsec:Slice-Resource-Demand}.

Each InP, considering the various slice resource reservation requests
received during some time interval, tries to maximizes the number
of slices for which the reservation can be satisfied. Costs induced
by the variation with the time of the resource reservation request
are taken into account by the InP, see Section~\ref{subsec:Slice-preparation}.
A resource reservation request for a slice is considered satisfied
when \emph{i) }enough resources are available to meet a target resource
requirements and \emph{ii) }the Impact Probability (IP) on other best-effort
services running on the infrastructure network remains below some
threshold $\overline{p}^{\text{im}}$, see Section~\ref{dy:subsec:Background-Services}.
The slice priority level is taken into account when processing the
reservation requests. The InP answers positively or negatively to
a reservation request. In the latter case, the MNO may contact alternative
InPs, or, when no InP has enough available resources, the MNO may
reject the slice service booking request from the SP or ask the SP
to update its slice service request (SM-SLA negotiation).

Table~\ref{dy:tab:notations} summarizes the main notations introduced
in this paper.\vspace{-0.5cm}

\begin{center}
\begin{table}[tbh]
\caption{Table of Main Notations\label{dy:tab:notations}}

\centering{}%
\begin{tabular}{cl}
\toprule 
\multicolumn{2}{l}{\textsc{Infrastructure network}}\tabularnewline
\midrule
$\mathcal{G}$ & Infrastructure network graph, $\mathcal{G}=\left(\mathcal{N},\mathcal{E}\right)$\tabularnewline
$\mathcal{N}$ & Set of infrastructure nodes\tabularnewline
$\mathcal{E}$ & Set of infrastructure links\tabularnewline
$\Upsilon$ & Set of node resource types, $\Upsilon=\left\{ \text{c},\text{m},\text{w}\right\} $\tabularnewline
$a_{n}\left(i\right)$ & Available resource of type $n\in\Upsilon$ at node~$i$\tabularnewline
$a_{\text{b}}\left(ij\right)$ & Available bandwidth of link~$ij$\tabularnewline
$c_{n}\left(i\right)$ & Per-unit cost of resource of type $n\in\Upsilon$ for node~$i$\tabularnewline
$c_{\text{b}}\left(ij\right)$ & Per-unit cost for link~$ij$\tabularnewline
$c_{\text{f}}\left(i\right)$ & Fixed cost for using node~$i$\tabularnewline
$c_{\text{a}}\left(i\right)$ & Reservation adaptation cost at node~$i$\tabularnewline
\midrule 
\multicolumn{2}{l}{\textsc{Temporal notations}}\tabularnewline
\midrule
$\mathcal{P}_{k}$ & Processing time interval in time slot $k$\tabularnewline
$T$ & Duration of a time slot\tabularnewline
$\varepsilon T$ & Processing duration (of $\mathcal{P}_{k}$)\tabularnewline
\midrule 
\multicolumn{2}{l}{\textsc{Slice requests and resource demands}}\tabularnewline
\midrule
$\mathcal{G}_{s}$ & SFC graph of slice $s$, $\mathcal{G}_{s}=\left(\mathcal{N}_{s},\mathcal{E}_{s}\right)$\tabularnewline
$\mathcal{N}_{s}$ & Set of virtual network functions\tabularnewline
$\mathcal{E}_{s}$ & Set of virtual links\tabularnewline
$P_{s}^{\text{c}}$ & Priority class\tabularnewline
$P_{s,k}$ & Priority level at time $k$\tabularnewline
$\mathcal{K}_{s}$ & Slice active interval, $\mathcal{K}_{s}=\left[k_{s}^{\text{on}},k_{s}^{\text{off}}\right]$\tabularnewline
$\boldsymbol{r}_{s}$ & Vector of resource demands of an SFC\tabularnewline
$\boldsymbol{U}_{s,k}$ & Vector of resource demands of a typical\tabularnewline
 & user in time slot $k$\tabularnewline
$\boldsymbol{R}_{s,k}$ & Vector of aggregate resource demands\tabularnewline
 & in time slot $k$\tabularnewline
$\boldsymbol{B}_{k}$ & Vector of resources consumed by background\tabularnewline
 & services in time slot $k$\tabularnewline
$\overline{U}$, $\overline{R}$, $\overline{B}$ & Mean value of $\boldsymbol{U}$, $\boldsymbol{R}$, $\boldsymbol{B}$\tabularnewline
$\widetilde{U}$, $\widetilde{R}$, $\widetilde{B}$ & Standard deviation of $\boldsymbol{U}$, $\boldsymbol{R}$, $\boldsymbol{B}$\tabularnewline
$\underline{p}_{s}^{\text{sp}}$ & Required service satisfaction probability\tabularnewline
$\overline{p}^{\text{im}}$ & Slice impact probability threshold\tabularnewline
 & (w.r.t. background services)\tabularnewline
$\mathcal{S}_{k}$ & Slices requests received before $\left(k+1\right)T-\varepsilon T$\tabularnewline
$\mathcal{R}_{k}$ & Slices requests processed during $\mathcal{P}_{k}$\tabularnewline
\bottomrule
\end{tabular}
\end{table}
\vspace{-0.5cm}
\par\end{center}

\subsection{Network Model\label{dy:subsec:Network-Model}}

In this paper, to simplify presentation, one considers an infrastructure
network owned by a single InP. The infrastructure network managed
by the considered InP is represented by a directed graph $\mathcal{G}=\left(\mathcal{N},\mathcal{E}\right)$.
$\mathcal{N}$ is the set of infrastructure nodes and $\mathcal{E}$
is the set of infrastructure links, which correspond to the wired
connections between and within nodes (loop-back links) of the infrastructure
network.

Each infrastructure node $i\in\mathcal{N}$ is characterized by its
computing $a_{\text{c}}\left(i\right)$, memory $a_{\text{m}}\left(i\right)$,
and wireless $a_{\text{w}}\left(i\right)$ resources. For each node
$i$, the InP charges the MNO a fixed cost $c_{\text{f}}\left(i\right)$
for node disposal (paid for each slice using node~$i$), and per-unit
variable costs $c_{n}(i)$, $n\in\Upsilon=\left\{ \text{c},\text{m},\text{w}\right\} $,
which depend linearly on the amount of resources provided by that
node.

Similarly, each infrastructure link $ij\in\mathcal{E}$ connecting
node $i$ to $j$ is characterized by its bandwidth $a_{\text{b}}\left(ij\right)$,
and an associated per-unit bandwidth cost $c_{\text{b}}(ij)$. Several
distinct VNFs of the same slice may be deployed on a given infrastructure
node. When communication between these VNFs is required, an internal
(loop-back) infrastructure link $ii\in\mathcal{E}$ can be used at
each node $i\in\mathcal{N}$, as in \cite{Wang2009}, in the case
of interconnected virtual machines (VMs) deployed on the same host.
In that case, the InP charges the MNO per-unit bandwidth cost $c_{\textrm{\text{b}}}\left(ii\right)$.

\subsection{Slice Resource Reservation Requests and Adaptation Costs\label{dy:subsec:Slice-Prov-Request}}

\subsubsection{Request Arrivals\label{dy:subsec:Request-Arrivals}}

One considers that time is slotted into slots of constant duration~$T$
(typically of few tens of minutes), which represents the time unit
considered for the slice resource reservation duration in the booking
calendar. The slot of index $k\in\mathbb{N}$ lasts over the time
interval $\left[kT,\left(k+1\right)T\right[$. One considers that
the slice lifetime spans over one or several time slots of duration
$T$. Resources have to be reserved so as to be compliant with the
variations of the number of users and of their demands during the
slice lifetime. The service characteristics are assumed stable over
each time slot, and may vary from one time slot to the next.

\begin{figure}[tbh]
\begin{centering}
\includegraphics[width=1\columnwidth]{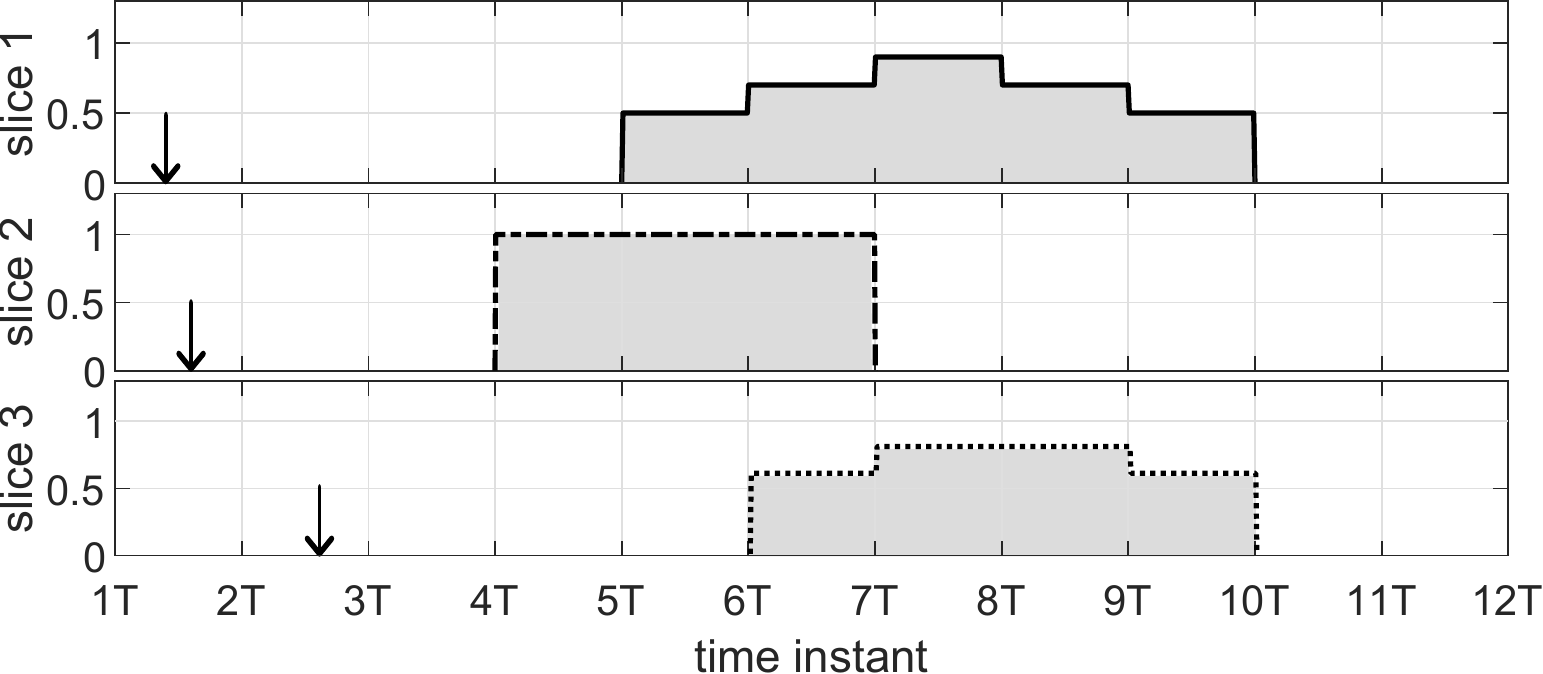}
\par\end{centering}
\caption{Arrivals of slice resource reservation requests as a function of time;
Black arrows represent the arrival times $t_{s}$ of each request;
The types of slices are illustrated by different plot line styles;
The slice resource demands evolve with time; Peak demands have been
normalized. \label{dy:fig:Intro:Prov_Interval}}
\end{figure}

Let $t_{s}$ be the time instant at which the reservation request
for a slice $s$ is received by the InP. This slice is also characterized
by the index $k_{s}^{\text{on}}$ of the time slot at the beginning
of which it has to be activated (put into service), and the index
$k_{s}^{\text{off}}$ of the time slot at the end of which it has
to be deactivated. Thus, the slice $s$ is active over the time interval
$\left[k_{s}^{\text{on}}T,\left(k_{s}^{\text{off}}+1\right)T\right[$.
Figure~\ref{dy:fig:Intro:Prov_Interval} depicts an example of arrivals
of slice resource reservation requests, as well as the time slots
over which the corresponding services have to be active.

\subsubsection{Slice Resource Demand\label{dy:subsec:Slice-Resource-Demand}}

A demand for resources for a slice $s$ is defined on the basis of
the translation of the SM-SLA between an SP and an MNO. A priority
class $P_{s}^{\text{c}}$ determines the priority level with which
the resource demand has to be processed. As in \cite{Luu:cov}, the
type of service provided by slice~$s$ is used to identify the type
of SFC, \textit{i.e}., the ordered set of VNFs involved in $s$. We
consider that a slice is devoted to a single type of service supplied
by a given type of SFC. Several SFCs of the same type may have to
be deployed so as to satisfy user demands within the slice. The topology
of each SFC of slice~$s$ is represented by a graph $\mathcal{G}_{s}=\left(\mathcal{N}_{s},\mathcal{E}_{s}\right)$
representing the VNFs and their interconnections. Each virtual node
$v\in\mathcal{N}_{s}$ represents a VNF, and each virtual link $vw\in\mathcal{E}_{s}$
represents the connection between virtual nodes~$v$ and $w$.

Based on $\mathcal{G}_{s}$, one introduces the vectors $\boldsymbol{r}_{s}$,
$\mathbf{U}_{s,k}$, and $\mathbf{R}_{s,k}$, respectively representing
the \emph{resource demands} of a single SFC (SFC-RD), of a typical
user (U-RD) during time slot~$k$, and the aggregate resource demand
of the users of slice~$s$ (S-RD) during time slot~$k$.

The deterministic SFC-RD vector
\begin{equation}
\boldsymbol{r}_{s}=\left[r_{s,n}\left(v\right),r_{s,\text{b}}\left(vw\right)\right]_{n\in\Upsilon,\left(v,vw\right)\in\mathcal{G}_{s}}^{\top}
\end{equation}
gathers the computing ($r_{s,\text{c}}\left(v\right)$), memory ($r_{s,\text{m}}\left(v\right)$),
wireless ($r_{s,\text{w}}\left(v\right)$), and bandwidth ($r_{s,\text{b}}\left(vw\right)$)
resource requirements of the VNFs $v\in\mathcal{N}_{s}$ and the virtual
links~$vw\in\mathcal{E}_{s}$ of a single SFC. The vector $\boldsymbol{r}_{s}$
is assumed to be time invariant, as it characterizes the resources
which need to be allocated to run an instance of the considered SFC.
In the considered reservation context, $\boldsymbol{r}_{s}$ also
represents the \emph{maximum} amount of reserved resources that will
be made available to the considered SFC.

Each user of slice~$s$ is assumed to consume a random amount of
the resources of an SFC of that slice. The random vector 
\begin{equation}
\mathbf{U}_{s,k}=\left[U_{s,n,k}\left(v\right),U_{s,\text{b},k}\left(vw\right)\right]_{n\in\Upsilon,\left(v,vw\right)\in\mathcal{G}_{s}}^{\top}
\end{equation}
of U-RD represents the resource demands of a single user of slice~$s$
during time slot $k$. $U_{s,n,k}\left(v\right)$, $n\in\Upsilon$,
and $U_{s,\text{b},k}\left(vw\right)$ are the random amounts of employed
resources of VNF~$v$ and of virtual link $vw$. In addition, the
resources consumed by various users are represented by independently
and identically distributed random vectors. Minor variations of the
user resource demand within time slot~$k$ are accounted for by the
probability distribution characterizing $\mathbf{U}_{s,k}$.

The random S-RD vector \textit{\emph{
\begin{equation}
\mathbf{R}_{s,k}=\left[R_{s,n,k}\left(v\right),R_{s,\text{b},k}\left(vw\right)\right]_{n\in\Upsilon,\left(v,vw\right)\in\mathcal{G}_{s}}^{\top}
\end{equation}
gathers }}$R_{s,n,k}\left(v\right)$, $n\in\Upsilon$, and $R_{s,\text{b},k}\left(vw\right)$,\textit{\emph{
}}the aggregate amount of resources employed by a random number $N_{s,k}$
of independent users of slice~$s$ during time slot~$k$. Minor
variations of the number of users within time slot~$k$ are as well
captured by the probability mass function characterizing $N_{s,k}$.

The probability distributions characterizing $\mathbf{U}_{s,k}$,
$N_{s,k}$, and consequently $\mathbf{R}_{s,k}$ depend on the time
slot index $k$, to represent possible large changes in the U-RD or
in the number of users of slice~$s$ in successive time slots.

One considers, for a typical user and during a given time slot~$k$,
that for each virtual node $v\in\mathcal{N}_{s}$, the resource demands
of different types $n\in\Upsilon$ are correlated. A correlation may
also exist between the demands for resources of the same type among
virtual nodes. Finally, the resulting traffic demands between nodes
is usually also correlated with the resource demands for a given virtual
node, as reported in \cite{Jiang2019}. Considering the U-RD vector
$\mathbf{U}_{s,k}$, one assumes that the elements $U_{s,n,k}\left(v\right)$,
$\forall n\in\Upsilon$, and $U_{s,\text{b},k}\left(vw\right)$ are
normally distributed during the time slot~$k$. $\mathbf{U}_{s,k}$
thus follows a multivariate normal distribution with probability density
\begin{align}
 & f\left(\mathbf{x};\boldsymbol{\mu}_{s,k},\boldsymbol{\Gamma}_{s,k}\right)=\nonumber \\
 & \quad\left(2\pi\right)^{-\frac{1}{2}\textrm{card}\left(\mathbf{U}_{s,k}\right)}\left|\boldsymbol{\Gamma}_{s,k}\right|^{-\frac{1}{2}}e^{-\frac{1}{2}\left(\mathbf{x}-\boldsymbol{\mu}_{s,k}\right)^{\top}\boldsymbol{\Gamma}_{s,k}^{-1}\left(\mathbf{x}-\boldsymbol{\mu}_{s,k}\right)},\label{dy:eq:JointResDist}
\end{align}
where $\textrm{card}\left(\mathbf{U}_{s,k}\right)$ is the number
of elements of $\mathbf{U}_{s,k}$, 
\begin{equation}
\boldsymbol{\mu}_{s,k}=\left[\overline{U}_{s,n,k}\left(v\right),\overline{U}_{s,\text{b},k}\left(vw\right)\right]_{\left(v,vw\right)\in\mathcal{G}_{s},n\in\Upsilon}^{\top}
\end{equation}
is its mean value and and $\boldsymbol{\Gamma}_{s,k}$ is its covariance
matrix, with diagonal elements
\begin{equation}
\text{diag}\left(\boldsymbol{\Gamma}_{s,k}\right)=\left[\widetilde{U}_{s,n,k}^{2}\left(v\right),\widetilde{U}_{s,\text{b},k}^{2}\left(vw\right)\right]_{\left(v,vw\right)\in\mathcal{G}_{s},n\in\Upsilon}^{\top},
\end{equation}
and off-diagonal elements representing the correlation between different
types of resource demands. One has thus 
\begin{align}
U_{s,n,k}(v)\sim & \mathcal{N}\left(\overline{U}_{s,n,k}\left(v\right),\widetilde{U}_{s,n,k}^{2}\left(v\right)\right),n\in\Upsilon\text{, and}\nonumber \\
U_{s,\text{b},k}\left(vw\right)\sim & \mathcal{N}\left(\overline{U}_{s,\text{b},k}\left(vw\right),\widetilde{U}_{s,\text{b},k}^{2}\left(vw\right)\right).
\end{align}

The probability that the number of users $N_{s,k}$ to be supported
by slice~$s$ in the $k$-th time slot is equal to $\eta$ is 
\begin{equation}
p_{s,k,\eta}=\Pr\left(N_{s,k}=\eta\right),\text{\ensuremath{\eta\in\mathbb{N}}}.\label{dy:eq:UserDist}
\end{equation}
The amount of resources of the VNF~$v$ and of the virtual link~$vw$
consumed by different users is represented by independently and identically
distributed copies of $\mathbf{U}_{s,k}$. Consequently, the joint
distribution of the aggregate amount $\mathbf{U}_{s,\eta,k}$ of resources
consumed by $\eta$ independent users is $f\left(\mathbf{x},\eta\boldsymbol{\mu}_{s,k},\eta^{2}\boldsymbol{\Gamma}_{s,k}\right)$.
The total amount of resources employed by a random number $N_{s,k}$
of independent users, $\mathbf{R}_{s,k}=\mathbf{U}_{s,N_{s,k},k}=\left(R_{s,n,k}\left(v\right),R_{s,\text{b},k}\left(vw\right)\right)_{n\in\Upsilon,\left(v,vw\right)\in\mathcal{G}_{s}}^{\top}$
, is distributed according to
\begin{equation}
g\left(\mathbf{x},\boldsymbol{\mu}_{s,k},\boldsymbol{\Gamma}_{s,k}\right)=\sum_{\eta=0}^{\infty}p_{s,k,\eta}f\left(\mathbf{x},\eta\boldsymbol{\mu}_{s,k},\eta^{2}\boldsymbol{\Gamma}_{s,k}\right).\label{dy:eq:PDF_Rs}
\end{equation}

\subsubsection{Adaptation Costs to Request Variations\label{subsec:Slice-preparation}}

During the lifetime of a slice, the amount of required slice resources
may evolve from one time slot to another. An increase of the required
resources may impact the resource allocation scheme by requiring more
infrastructure resources to be allocated. Compared to a situation
where the resource allocation is static for the whole lifespan of
a slice, this induces more operations to be performed on the network
infrastructure (assignment or re-assignment of resources, launching
virtual machines or containers on which VNFs will be operated) and
results in additional costs to the InP. The anticipation of those
allocation adaptation costs are then considered when processing the
slice resource reservation request. A cost $c_{\textrm{a}}\left(i\right)$
for each unit \emph{increase} of the amount of instances (\emph{i.e.},
virtual machines or containers) of a VNF between two time slots is
assumed to be charged by the InP to the MNO. Resource release costs
are assumed to be incorporated within $c_{\textrm{a}}\left(i\right)$.

As will be seen in Section~\ref{dy:subsec:Costs-and-incomes}, this
cost reduces SFC migrations within a given slice between consecutive
time slots.

\subsection{Resource Consumption of Background Services\label{dy:subsec:Background-Services}}

In a given time slot $k$, we assume that infrastructure resources
are partly consumed by best-effort background services for which no
resource reservation has been performed. One denotes
\begin{equation}
\mathbf{B}_{k}=\left[B_{n,k}\left(i\right),B_{\text{b},k}\left(ij\right)\right]_{\left(i,ij\right)\in\mathcal{G},n\in\Upsilon}^{\top}
\end{equation}
the vector gathering all resources consumed by background services
during time slot~$k$. The elements $B_{\text{c},k}\left(i\right)$,
$B_{\text{m},k}\left(i\right)$, $B_{\text{w},k}\left(i\right)$,
$\forall i\in\mathcal{N}$, and $B_{\text{b},k}\left(ij\right)$,
$\forall ij\in\mathcal{E}$ of $\mathbf{B}_{k}$ are random variables
representing the aggregate amount of computing, memory, wireless,
and bandwidth resources consumed by these best-effort services. As
in \cite{Luu:unc}, each of those variables is assumed to be uncorrelated
and Gaussian distributed, 
\begin{align}
B_{n,k}\left(i\right)\sim & \mathcal{N}\left(\overline{B}_{n,k}\left(i\right),\widetilde{B}_{n,k}^{2}\left(i\right)\right),\forall i\in\mathcal{N},\forall n\in\Upsilon\text{, and}\nonumber \\
B_{\text{b},k}\left(ij\right)\sim & \mathcal{N}\left(\overline{B}_{\text{b},k}\left(ij\right),\widetilde{B}_{\text{b},k}^{2}\left(ij\right)\right),\forall ij\in\mathcal{E}.
\end{align}
Consequently, during each time slot $k$, $\mathbf{B}_{k}$ is distributed
according to $f\left(\mathbf{x};\boldsymbol{\mu}_{\text{B},k},\boldsymbol{\Gamma}_{\text{B},k}\right)$,
with
\begin{align}
\boldsymbol{\mu}_{\text{B},k} & =\left[\overline{B}_{n,k}\left(i\right),\overline{B}_{\text{b},k}\left(ij\right)\right]_{\left(i,ij\right)\in\mathcal{G},n\in\Upsilon}^{\top},\\
\boldsymbol{\Gamma}_{\text{B},k} & =\text{diag}\left[\widetilde{B}_{n,k}^{2}\left(i\right),\widetilde{B}_{\text{b},k}^{2}\left(ij\right)\right]_{\left(i,ij\right)\in\mathcal{G},n\in\Upsilon}.
\end{align}
The evolution of resources consumed by background services over time
slots may be predicted by the InP from past observations, see, \emph{e.g.},
\cite{Tan2011}. The smaller variations within each time slot are
taken into account in the probability distribution.

\section{Slice Resource Reservation Approaches\label{dy:sec:Reservation-Approaches}}

\begin{figure}[tbh]
\begin{centering}
\includegraphics[width=0.6\columnwidth]{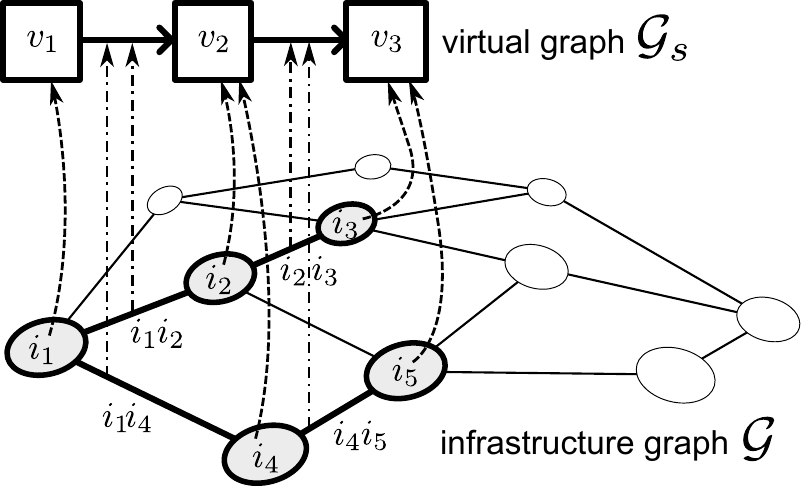}
\par\end{centering}
\caption{Reservation of infrastructure resources for slice~$s$: Resources
from the infrastructure node~$i_{1}$ and the aggregate resources
from the infrastructure node pairs~$\left(i_{2},i_{4}\right)$ and
$\left(i_{3},i_{5}\right)$ are respectively reserved for the virtual
nodes~$v_{1}$, $v_{2}$, and $v_{3}$. Correspondingly, resources
of the infrastructure link pairs $\left(i_{1}i_{2},i_{1}i_{4}\right)$
and $\left(i_{2}i_{3},i_{4}i_{5}\right)$ (highlighted by the bold
lines) are reserved for the virtual links $v_{1}v_{2}$ and $v_{2}v_{3}$.
\label{dy:fig:Intro_Mapping}}
\end{figure}

Taking the InP perspective, slice resource reservation aims at booking,
somewhat in advance, enough infrastructure resources to ensure that
the MNOs will be able to provide slices with characteristics as stated
in the SM-SLA. For that purpose, the InP has to identify \emph{i})\emph{
}the infrastructure nodes which will provide resources for the future
deployment of VNFs and \emph{ii}) the links able to transmit data
between these nodes/VNFs, while respecting the structure of SFCs.
This correspond to the network environment preparation block represented
in Figure~\ref{dy:fig:Intro:SliceLifeCycle}. Within some time slot
over which a slice $s$ is active, the slice resource reservation
can be represented by a mapping between the infrastructure graph $\mathcal{G}$
and the S-RD graph $\mathcal{G}_{s}$ as illustrated in Figure~\ref{dy:fig:Intro_Mapping}.
In this example, the slice~$s$ consists of several linear SFCs of
the same type for which resources have been reserved from some infrastructure
network.

The mapping between $\mathcal{G}$ and $\mathcal{G}_{s}$ may evolve
between successive time intervals due to the evolution of the characteristics
of the MI-SLA for slice $s$, to the arrival of new slice resource
reservation requests, and to resources released by terminated slices.

\subsection{Prioritized Processing of Resource Reservation Requests\label{dy:subsec:Provisioning-Scheduling}}

Several strategies for processing resource reservation requests can
be considered to account for their dynamicity. A first approach consists
in processing the resource reservation requests as soon as they are
submitted by an MNO. The advantage is to immediately indicate to the
MNO whether enough infrastructure resources are available to satisfy
the request. A second approach is to wait some time and process several
requests simultaneously. This second approach, considered in this
paper, helps process the resource reservation requests since the InP
has a better view of concurrent requests. Additionally, by a slice-priority
differentiated processing delay of the slice reservation request a
compromise between response delay to the MNO and optimization of the
reservation of resources by the InP can be found.

When processing a new request, already granted resources requests
may be adjusted. This update possibility gives more degrees of freedom
to the InP to satisfy new requests, but comes at the price of higher
computational complexity. Updates must be done while satisfying previous
requests which have been indicated to the MNOs as granted. In this
paper, we have chosen not to change any assignment of previously successfully
processed slice requests.

Independently of the chosen strategy, the InP has to account for the
time required for processing the resource reservation\textit{\emph{
and performing the }}slice deployment and activation (lasting few
minutes, as indicated in \cite{Boubendir2018}). Consequently, resource
reservation requests for slices to be activated at $\left(k+1\right)T$
should reach the InP before $\left(k+1\right)T-\varepsilon T$, where
$\varepsilon T$, $\varepsilon\in\left]0,1\right[$, is an upper bound
of the time required for the slice resource request processing operations,
the slice activation and updates.

Let $\mathcal{S}_{k}$ be the set of slices whose resource reservation
requests have been received before $\left(k+1\right)T-\varepsilon T$.
A flag $f_{s}\in\left\{ 0,1\right\} $ indicates for each slice $s\in\mathcal{S}_{k}$
whether the request has been processed ($f_{s}=1$) (granted or denied)
or is still to be processed ($f_{s}=0$).

In what follows, we consider two classes of slices, namely Premium
and Standard. The priority level is indicated by the MNO to the InP
in the MI-SLA of the slice. Each slice request, when received for
the first time in the interval $\mathcal{T}_{k}=\left[kT-\varepsilon T,\left(k+1\right)T-\varepsilon T\right[$,
gets $f_{s}=0$, and is assigned a priority level $P_{s,k}\in\mathbb{R}$
depending on its class
\begin{equation}
P_{s,k}=\begin{cases}
P_{\max} & \text{for Premium slices,}\\
0 & \text{for Standard slice, if }k_{s}^{\text{on}}>k+1,\\
P_{\max}-1 & \text{for Standard slice, if }k_{s}^{\text{on}}=k+1.
\end{cases}\hspace{-0.158cm}\label{dy:eq:AssignPriority}
\end{equation}
Standard slice requests, which have to be activated in the next time
slot, get thus a higher priority level. Then, only slices whose priority
level is above a certain threshold
\begin{equation}
P_{\text{thres}}=\alpha\left(P_{\max}-1\right),\,\text{with }\alpha\in\left[0,1\right],
\end{equation}
are processed in the time interval $\mathcal{P}_{k}=\left[\left(k+1\right)T-\varepsilon T,\left(k+1\right)T\right[$
of duration~$\varepsilon T$. The set of slices whose resource request
has to be processed during the time interval $\mathcal{P}_{k}$ is
\begin{equation}
\mathcal{R}_{k}\triangleq\left\{ s\in\mathcal{S}_{k}:f_{s}=0,P_{s,k}\geqslant P_{\text{thres}}\right\} .
\end{equation}
Once the resource request of a slice in $\mathcal{R}_{k}$ is processed,
its flag is set to $f_{s}=1$. All standard slice requests with $P_{s,k}<P_{\text{thres}}$
(pending requests) are delayed and may be processed in the next time
interval $\mathcal{P}_{k+1}$. Their priority is updated as 
\begin{align}
P_{s,k+1} & =\begin{cases}
\min\left\{ P_{s,k}+\Delta P,P_{\max}-1\right\}  & \text{if }k_{s}^{\text{on}}>k+2,\\
P_{\max}-1 & \text{if }k_{s}^{\text{on}}=k+2,
\end{cases}
\end{align}
where $\Delta P\geqslant0$ is some priority increment. When several
slices of equal priority have to be processed in a given time slot,
a possible choice, adopted in this paper, is to process first those
who have to be activated first, then those who have been submitted
first. Premium slices are always processed first. The processing delay
of Standard slice requests depends thus on $\alpha$ and $\Delta P$.
Deferring more the processing of Standard slice requests gives more
chance to satisfy Premium slice requests.

When $\alpha=0$, whatever the value of $\Delta P$, all slices resource
reservation requests received in the time interval $\mathcal{T}_{k}$
are processed, starting from the Premium slices, with the risk of
having no resources available for Premium slice requests received
in the few next time slots. This corresponds to the as-they-arrive
processing approach, considered, \emph{e.g.}, in \cite{Salvat2018}.

When $\alpha=1$ and $\Delta P=0$, the processing of Standard slice
resource reservation requests is delayed until the time slot preceding
their activation, leaving a maximum amount of resources available
for Premium slice resource reservation requests. Standard slice requests
are always processed \emph{just-in-time}, while processing of Premium
requests is anticipated.

Figure~\ref{dy:fig:Intro_TimeSlot} illustrates a scenario taking
place during the processing time interval $\mathcal{P}_{k}$ when
the processing of Standard slice requests is maximally delayed ($\alpha=1$
and $\Delta P=0$). The three slice requests $s_{1}$, $s_{2}$, and
$s_{3}$ in $\mathcal{S}_{k}$ are assumed still to be processed.
The slice request $s_{4}$ arrives within $\mathcal{P}_{k}$ and will
thus be considered in $\mathcal{P}_{k+1}$. Among the slices $s_{1},s_{2}$,
and $s_{3}$, only $s_{3}$ is Premium (the time instant at which
the resource reservation request is submitted is indicated by a solid
arrow), and is therefore processed in $\mathcal{P}_{k}$. The slice
requests $s_{1}$ and $s_{2}$ are Standard (reservation request time
instants indicated by dashed arrows). They have to be active in the
time slots $\mathcal{K}_{s_{1}}=\left[k_{s_{1}}^{\text{on}},k_{s_{1}}^{\text{off}}\right]$
and $\mathcal{K}_{s_{2}}=\left[k_{s_{2}}^{\text{on}},k_{s_{2}}^{\text{off}}\right]$.
Since $k_{s_{1}}^{\text{on}}=k+1$ and $k_{s_{2}}^{\text{on}}=k+2$,
only $s_{1}$ is processed in $\mathcal{P}_{k}$. Finally, the set
of slice resource reservation requests to be processed in $\mathcal{P}_{k}$
is $\mathcal{R}_{k}=\left\{ s_{1},s_{3}\right\} $ (highlighted by
red arrows).
\begin{center}
\begin{figure}[tbh]
\begin{centering}
\includegraphics[width=1\columnwidth]{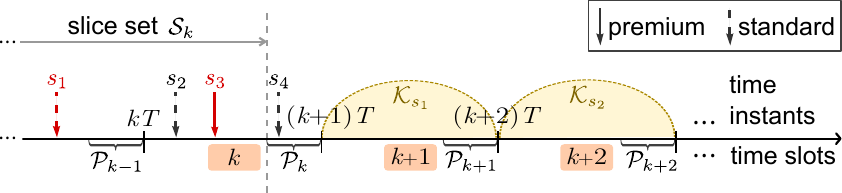}
\par\end{centering}
\caption{Time slots, arrival times of the slice resource reservation requests,
and time intervals during which the reservation is processed. \label{dy:fig:Intro_TimeSlot}}
\end{figure}
\vspace{-0.7cm}
\par\end{center}

\subsection{Decision Variables\label{dy:subsec:Decision-Variables}}

Processing a resource reservation request for some slice~$s\in\mathcal{R}_{k}$
amounts to defining a mapping $\boldsymbol{\kappa}_{s,\ell}$ between
the graphs $\mathcal{G}=\left(\mathcal{N},\mathcal{E}\right)$ and
$\mathcal{G}_{s}=\left(\mathcal{N}_{s},\mathcal{E}_{s}\right)$ for
each time slot $\ell\in\mathcal{K}_{s}\triangleq\left[k_{s}^{\text{on}},k_{s}^{\text{off}}\right]$
during which the slice~$s$ is active. This mapping describes \emph{i})\emph{
}the number $\kappa_{s,\ell}\left(i,v\right)\in\mathbb{N}$ of VNF
instances of type $v\in\mathcal{N}_{s}$ for which node~$i\in\mathcal{N}$
will reserve resources, and \emph{ii}) the number $\kappa_{s,\ell}\left(ij,vw\right)\in\mathbb{N}$
of links $vw\in\mathcal{E}_{s}$ between VNF instances for which the
InP will reserve resources on the infrastructure link $ij\in\mathcal{E}$,
both in time slot~$\ell$. The amount of resource of type $n\in\Upsilon$
reserved by node~$i$ for a VNF instance of type~$v$ is $\kappa_{s,\ell}\left(i,v\right)r_{s,n}(v)$.
The bandwidth reserved on link~$ij$ to support the traffic between
two virtual nodes of type~$v$ and $w$ is represented $\kappa_{s,\ell}\left(ij,vw\right)r_{s,\text{b}}(vw)$.

The mapping $\boldsymbol{\kappa}_{s,\ell}$ is thus defined as
\begin{equation}
\boldsymbol{\kappa}_{s,\ell}=\left\{ \hspace{-0.2cm}\begin{array}{l}
\kappa_{s,\ell}\left(i,v\right),\\
\kappa_{s,\ell}\left(ij,vw\right)
\end{array}\hspace{-0.2cm}\right\} _{\left(i,ij\right)\in\mathcal{G},\left(v,vw\right)\in\mathcal{G}_{s}}
\end{equation}
for each $\ell\in\mathcal{K}_{s}$. By convention, $\boldsymbol{\kappa}_{s,\ell}=\mathbf{0}$
when $\ell\notin\mathcal{K}_{s}$. Moreover, one introduces $\boldsymbol{\kappa}_{s}=\left\{ \boldsymbol{\kappa}_{s,\ell}:\ell\in\mathcal{K}_{s}\right\} $
to gather all assignments performed for the slice~$s$.

Enough infrastructure resources are not always available for a given
slice $s\in\mathcal{R}_{k}$. The binary decision variable $d_{s}$
indicates whether all conditions are met to satisfy the resource reservation
request for slice~$s$ and consequently whether resources are actually
reserved for slice $s$ ($d_{s}=1$) or not ($d_{s}=0$). These conditions
are detailed in the following sections.

Consequently, the set of variables which have to be assigned by the
InP in the processing time interval $\mathcal{P}_{k}$ are
\begin{align}
\mathbf{d}_{\mathcal{R}_{k}} & =\left\{ d_{s}:s\in\mathcal{R}_{k}\right\} ,\text{ and}\\
\boldsymbol{\kappa}_{\mathcal{R}_{k}} & =\left\{ \boldsymbol{\kappa}_{s,\ell}:s\in\mathcal{R}_{k},\ell\in\mathcal{K}_{s}\right\} .
\end{align}
The vector $\mathbf{d}_{\mathcal{R}_{k}}$ indicates which slice resource
reservation requests in $\mathcal{R}_{k}$ have been granted, and
$\boldsymbol{\kappa}_{k}$ describes the way resources have been reserved
by the InP.

\subsection{Constraints\label{dy:subsec:Constraints}}

During the processing time interval $\mathcal{P}_{k}$ of time slot~$k$,
the InP has to account for all resource reservation requests of slices
$s\in\mathcal{S}_{k\textrm{-}1}$ which have been previously processed,
\emph{i.e.}, with $f_{s}=1$. The set of these slices is denoted as
\begin{equation}
\mathcal{S}_{k\textrm{-}1}^{\text{p}}=\left\{ s\in\mathcal{S}_{k\textrm{-}1}:f_{s}=1\right\} .
\end{equation}
Moreover, the mappings $\boldsymbol{\kappa}_{s,\ell}$ for all slices
$s\in\mathcal{R}_{k}$ have to satisfy some constraints to ensure
that \textit{i}) enough resources are reserved to properly deploy
the SFCs and \textit{ii}) the SSP $\underline{p}_{s}^{\text{sp}}$
is reached. These constraints have to be satisfied for all time slots
during which the slice is active. The InP has also to keep the impact
probability on background services below $\overline{p}^{\text{im}}$.
These constraints are described in what follows.

The total amount of resources reserved (and allocated for slices in
service) by each infrastructure node~$i\in\mathcal{N}$ and each
infrastructure link~$ij\in\mathcal{E}$ for all slices~$s\in\mathcal{R}_{k}$
has to be less than their available resources, see Section~\ref{dy:subsec:Network-Model}.
Consequently, the following constraints have to be satisfied, for
each~$\ell=\min_{s\in\mathcal{R}_{k}}\left\{ k_{s}^{\text{on}}\right\} \geqslant k,\dots,\max_{s\in\mathcal{R}_{k}}\left\{ k_{s}^{\text{off}}\right\} $,
\begin{align}
 & \sum_{s\in\mathcal{R}_{k}}\sum_{v\in\mathcal{N}_{s}}\kappa_{s,\ell}\left(i,v\right)r_{s,n}\left(v\right)\leqslant a_{n}\left(i\right)-\nonumber \\
 & \hspace{1cm}\sum_{s\in\mathcal{S}_{k\textrm{-}1}^{\text{p}}}\sum_{v\in\mathcal{N}_{s}}\kappa_{s,\ell}\left(i,v\right)r_{s,n}\left(v\right),\forall i\in\mathcal{N},n\in\Upsilon,\label{dy:eq:Cplx:Cons:Limit:Node}
\end{align}
\begin{align}
 & \sum\limits _{s\in\mathcal{R}_{k}}\sum_{vw\in\mathcal{E}_{s}}\kappa_{s,\ell}\left(ij,vw\right)r_{s,\text{b}}\left(vw\right)\leqslant a_{\text{b}}\left(ij\right)-\nonumber \\
 & \hspace{1cm}\sum\limits _{s\in\mathcal{S}_{k\textrm{-}1}^{\text{p}}}\sum_{vw\in\mathcal{E}_{s}}\kappa_{s,\ell}\left(ij,vw\right)r_{s,\text{b}}\left(vw\right),\forall ij\in\mathcal{E}.\label{dy:eq:Cplx:Cons:Limit:Link}
\end{align}
In \eqref{dy:eq:Cplx:Cons:Limit:Node} and \eqref{dy:eq:Cplx:Cons:Limit:Link},
the right-hand sides of the inequalities represent the remaining part
of the resources once previous resource reservation requests have
been processed. When updates for granted slice requests are allowed,
$\boldsymbol{\kappa}_{s,\ell}$, $s\in\mathcal{S}_{k\textrm{-}1}^{\text{p}}$
are considered as variables, but not $d_{s}$, $s\in\mathcal{S}_{k\textrm{-}1}^{\text{p}}$,
since the status of successfully processed resource reservation requests
should not be changed. In what follows, one considers that such updates
are not allowed.

The inequalities \eqref{dy:eq:Cplx:Cons:Limit:Node} and \eqref{dy:eq:Cplx:Cons:Limit:Link}
may be more compactly written for ~$\ell>k$ as follows
\begin{align}
 & \sum_{s\in\mathcal{R}_{k}\cup\mathcal{S}_{k\textrm{-}1}^{\text{p}}}\sum_{v\in\mathcal{N}_{s}}\kappa_{s,\ell}\left(i,v\right)r_{s,n}\left(v\right)\leqslant a_{n}\left(i\right),\label{dy:eq:Cplx:Cons:Limit:Node-bis}\\
 & \sum\limits _{s\in\mathcal{R}_{k}\cup\mathcal{S}_{k\textrm{-}1}^{\text{p}}}\sum_{vw\in\mathcal{E}_{s}}\kappa_{s,\ell}\left(ij,vw\right)r_{s,\text{b}}\left(vw\right)\leqslant a_{\text{b}}\left(ij\right).\label{dy:eq:Cplx:Cons:Limit:Link-bis}
\end{align}
The conditions \eqref{dy:eq:Cplx:Cons:Limit:Node} and \eqref{dy:eq:Cplx:Cons:Limit:Node-bis}
are equivalent, as $\boldsymbol{\kappa}_{s,\ell}=\mathbf{0}$ when
$\ell\notin\mathcal{K}_{s}$. The same is also true for conditions
\eqref{dy:eq:Cplx:Cons:Limit:Link} and \eqref{dy:eq:Cplx:Cons:Limit:Link-bis}.

For each virtual link $vw\in\mathcal{E}_{s}$, resources on a sequence
of infrastructure links must be reserved between \textit{each} pair
of infrastructure nodes that have reserved resources to the virtual
nodes $v$ and $w$. This leads to the following flow conservation
constraint, for each $\ell\in\mathcal{K}_{s}$, $s\in\mathcal{R}_{k}$,
$i\in\mathcal{N}$, and $vw\in\mathcal{E}_{s}$, 
\begin{align}
 & \hspace{-0.1cm}\sum\limits _{j\in\mathcal{N}}\left[\kappa_{s,\ell}\left(ij,vw\right)-\kappa_{s,\ell}\left(ji,vw\right)\right]=\nonumber \\
 & \hspace{-0.1cm}\Bigg(\frac{r_{s,\text{b}}(vw)}{\sum\limits _{vu\in\mathcal{E}_{s}}r_{s,\text{b}}(vu)}\Bigg)\kappa_{s,\ell}\left(i,v\right)-\Bigg(\frac{r_{s,\text{b}}(vw)}{\sum\limits _{uw\in\mathcal{E}_{s}}r_{s,\text{b}}(uw)}\Bigg)\kappa_{s,\ell}\left(i,w\right),\label{dy:eq:Cplx:Cons:Flow}
\end{align}
which is similar to that introduced in \cite{Luu:cov}. More specifically,
\eqref{dy:eq:Cplx:Cons:Flow} imposes that the reserved bandwidth
is commensurate with the reserved node resources. This allows the
MNO to find an appropriate VNF embedding and chaining solution. Finally,
when the SFC embedding is performed, each SFC node will be mapped
onto one single node, and each SFC link will be mapped onto one single
path. SFCs are not allowed to be mapped towards multiple physical
paths.

\subsection{Demand Satisfaction and Impact Probabilities}

An assignment $\boldsymbol{\kappa}_{s,\ell}$, $\ell\in\mathcal{K}_{s}$
of a given slice $s\in\mathcal{R}_{k}$, which satisfies \eqref{dy:eq:Cplx:Cons:Limit:Node}\textendash \eqref{dy:eq:Cplx:Cons:Flow},
has to ensure a SSP above $\underline{p}_{s}^{\text{sp}}$ for all
time slots $\ell\in\mathcal{K}_{s}$ during which slice~$s$ is active.
As introduced in Section~\ref{sec:Problem-Statement}, the SSP for
a slice is the probability that resources reserved for the deployment
of the SFCs of that slice will meet the demand of users of the service.
This leads to the following conditions, for all $\ell\in\mathcal{K}_{s}$
and $s\in\mathcal{R}_{k}$,
\begin{align}
p_{s,\ell}\left(\boldsymbol{\kappa}_{s,\ell},d_{s}\right)\geqslant\underline{p}_{s}^{\text{sp}},\label{dy:eq:Cplx:Cons:Success}
\end{align}
where
\begin{align}
 & \hspace{-0.1cm}p_{s,\ell}\left(\boldsymbol{\kappa}_{s,\ell},d_{s}\right)=\nonumber \\
 & \hspace{-0.25cm}\begin{array}{cl}
\Pr\Big\{ & \hspace{-0.35cm}\sum\limits _{i\in\mathcal{N}}\kappa_{s,\ell}\left(i,v\right)r_{s,n}\left(v\right)\geqslant d_{s}R_{s,n,\ell}\left(v\right),\forall v\in\mathcal{N}_{s},n\in\Upsilon,\\
 & \hspace{-0.35cm}\sum\limits _{ij\in\mathcal{E}}\kappa_{s,\ell}\left(ij,vw\right)r_{s,\text{b}}\left(vw\right)\geqslant d_{s}R_{s,\text{b},\ell}\left(vw\right),\forall vw\in\mathcal{E}_{s}\Big\}.
\end{array}\hspace{-0.2cm}\label{dy:eq:Cplx:Proba:Success}
\end{align}
The variable $d_{s}$ in \eqref{dy:eq:Cplx:Proba:Success} is introduced
to cancel the SSP constraint when the request for a slice $s\in\mathcal{R}_{k}$
is not granted ($d_{s}=0$). In this case, as described later in Section~\ref{dy:subsec:Costs-and-incomes},
the minimization of the cost function yields $\boldsymbol{\kappa}_{s,\ell}=\mathbf{0}$,
$\forall\ell\in\mathcal{K}_{s}$, which satisfies \eqref{dy:eq:Cplx:Proba:Success},
thus avoiding the reservation of any resource. The evaluation of \eqref{dy:eq:Cplx:Proba:Success}
is detailed in Section~\ref{dy:sec:Appendix_IP}.

The constraints~\eqref{dy:eq:Cplx:Cons:Limit:Node} and \eqref{dy:eq:Cplx:Cons:Limit:Link}
ensure that the resources reserved for the slices~$s\in\mathcal{R}_{k}$
are less than the available resources in each infrastructure node~$i\in\mathcal{N}$
and infrastructure link~$ij\in\mathcal{E}$. Nevertheless, for some
$s\in\mathcal{R}_{k}$, the assignments $\boldsymbol{\kappa}_{s,\ell}$,
$\ell\in\mathcal{K}_{s}$ evaluated in the processing time interval
$\mathcal{P}_{k}$, taking into account all previously processed reservation
requests (described by $\boldsymbol{\kappa}_{\mathcal{S}_{k-1}}=\{\boldsymbol{\kappa}_{s,\ell}:s\in\mathcal{S}_{k-1},\ell\in\mathcal{K}_{s}\}$),
may be such that in time slot $\ell\geqslant k_{s}^{\text{on}}$,
not enough resources are left for the background best-effort services
described in Section~\ref{dy:subsec:Background-Services}, and may
then significantly affect such services. Consequently, in the processing
time interval $\mathcal{P}_{k}$, when evaluating $\boldsymbol{\kappa}_{\mathcal{R}_{k}}$,
one should have, for all $\ell>k$,
\begin{align}
p_{n,\ell}^{\text{im}}\left(\boldsymbol{\kappa}_{\mathcal{R}_{k}},i\right) & \leqslant\overline{p}^{\text{im}},\forall n\in\Upsilon,\forall i\in\mathcal{N},\label{dy:eq:Cplx:Cons:Impact:Node}\\
p_{\text{b},\ell}^{\text{im}}\left(\boldsymbol{\kappa}_{\mathcal{R}_{k}},ij\right) & \leqslant\overline{p}^{\text{im}},\forall ij\in\mathcal{E},\label{dy:eq:Cplx:Cons:Impact:Link}
\end{align}
where
\begin{align}
\hspace{-0.18cm}p_{n,\ell}^{\text{im}}\left(\boldsymbol{\kappa}_{\mathcal{R}_{k}},i\right)={\normalcolor \textrm{Pr}}\,\Big\{ & \sum\limits _{s\in\mathcal{R}_{k}\cup\mathcal{S}_{k\textrm{-}1}^{\text{p}}}\sum_{v\in\mathcal{N}_{s}}\kappa_{s,\ell}\left(i,v\right)r_{s,n}\left(v\right)\nonumber \\
 & \quad\geqslant a_{n}\left(i\right)-B_{n,\ell}\left(i\right)\Big\},\label{dy:eq:Cplx:Proba:Impact:Node}\\
\hspace{-0.18cm}p_{\text{b},\ell}^{\text{im}}\left(\boldsymbol{\kappa}_{\mathcal{R}_{k}},ij\right)={\normalcolor \textrm{Pr}}\,\Big\{ & \sum\limits _{s\in\mathcal{R}_{k}\cup\mathcal{S}_{k\textrm{-}1}^{\text{p}}}\sum_{vw\in\mathcal{E}_{s}}\kappa_{s,\ell}\left(ij,vw\right)r_{s,\text{b}}\left(vw\right)\nonumber \\
 & \quad\geqslant a_{\text{b}}\left(ij\right)-B_{\text{b},\ell}\left(ij\right)\Big\}.\label{dy:eq:Cplx:Proba:Impact:Link}
\end{align}
As indicated before, the evaluations of \eqref{dy:eq:Cplx:Proba:Impact:Node}
and \eqref{dy:eq:Cplx:Proba:Impact:Link} for all $\ell>k$ involve
$\boldsymbol{\kappa}_{\mathcal{S}_{k\textrm{-}1}^{\text{p}}}$ which
have already been evaluated in previous processing time intervals
and are considered as constants in the current processing time interval,
\textit{i.e}., $\mathcal{P}_{k}$. The dependency in $\boldsymbol{\kappa}_{\mathcal{S}_{k\textrm{-}1}^{\text{p}}}$
of $p_{n,\ell}^{\text{im}}$ and $p_{\text{b},\ell}^{\text{im}}$
is omitted to lighten notations.

The constraints \eqref{dy:eq:Cplx:Cons:Impact:Node} ensure that the
reserved resources have a limited impact on background services at
each node $i\in\mathcal{N}$. The constraints \eqref{dy:eq:Cplx:Cons:Impact:Link}
have the same role for the infrastructure links. The value of $\overline{p}^{\text{im}}$
is chosen by the InP to provide sufficient resources for the background
services at every infrastructure nodes and links. A small value of
$\overline{p}^{\text{im}}$ leads to a small impact of reserved resources
for slices on background services, but makes the resource reservation
problem more difficult compared to the case of $\overline{p}^{\text{im}}$
close to one, where the impact on background service is less taken
into account. The InP, by adjusting the impact probability threshold
$\overline{p}^{\text{im}}$, can trade the revenues provided by slices
and those provided by the background services. For example, choosing
$\overline{p}^{\text{im}}=1$ may leave no resources for background
services.

\subsection{Relaxation of Probabilistic Constraints\label{dy:subsec:Relax:Proba}}

This section introduces the relaxation of the probabilistic constraints
\eqref{dy:eq:Cplx:Cons:Success}, \eqref{dy:eq:Cplx:Cons:Impact:Node},
and \eqref{dy:eq:Cplx:Cons:Impact:Link}. These constraints are nonlinear
due to the need to evaluate $p_{s,\ell}\left(\boldsymbol{\kappa}_{s,\ell},d_{s}\right)$,
$p_{n,\ell}^{\text{im}}\left(\boldsymbol{\kappa}_{\mathcal{R}_{k}},i\right)$,
and $p_{\text{b},\ell}^{\text{im}}\left(\boldsymbol{\kappa}_{\mathcal{R}_{k}},ij\right)$
with \eqref{dy:eq:Cplx:Proba:Success}, \eqref{dy:eq:Cplx:Proba:Impact:Node},
and \eqref{dy:eq:Cplx:Proba:Impact:Link}. Using the approach introduced
in \cite{Luu:unc}, the MNO translates the SSP constraint~\eqref{dy:eq:Cplx:Cons:Success},
for all $s\in\mathcal{R}_{k}$ and $\ell\in\mathcal{K}_{s}$, into
the following linear deterministic constraints
\begin{align}
\hspace{-0.42cm}\sum\limits _{i\in\mathcal{N}}\kappa_{s,\ell}\left(i,v\right)r_{s,n}\left(v\right) & \geqslant d_{s}\widehat{R}_{s,n,\ell}\left(v,\gamma_{s,\ell}\right),\forall v,n,\label{dy:eq:Relax:Satisfy:Node}\\
\hspace{-0.42cm}\sum\limits _{ij\in\mathcal{E}}\kappa_{s,\ell}\left(ij,vw\right)r_{s,\text{b}}\left(vw\right) & \geqslant d_{s}\widehat{R}_{s,\text{b},\ell}\left(vw,\gamma_{s,\ell}\right),\forall vw,\label{dy:eq:Relax:Satisfy:Link}
\end{align}
where
\begin{align}
\widehat{R}_{s,n,\ell}\left(v,\gamma_{s,\ell}\right) & =\overline{R}_{s,n,\ell}\left(v\right)+\gamma_{s,\ell}\widetilde{R}_{s,n,\ell}\left(v\right),\label{dy:eq:MILP:Bound:Rn}\\
\widehat{R}_{s,\text{b},\ell}\left(vw,\gamma_{s,\ell}\right) & =\overline{R}_{s,\text{b},\ell}\left(vw\right)+\gamma_{s,\ell}\widetilde{R}_{s,\text{b},\ell}\left(vw\right),\label{dy:eq:MILP:Bound:Rb}
\end{align}
are the target aggregate user demands, depending on some parameter
$\gamma_{s,\ell}>0$. $\overline{R}_{s,n,\ell}\left(v\right)$ and
$\widetilde{R}_{s,n,\ell}\left(v\right)$ are the mean and standard
deviation of $R_{s,n,\ell}\left(v\right)$, while $\overline{R}_{s,\text{b},\ell}\left(vw\right)$
and $\widetilde{R}_{s,\text{b},\ell}\left(vw\right)$ are the mean
and standard deviation of $R_{s,\text{b},\ell}\left(vw\right)$. These
quantities are evaluated by the MNO. Appendix~\ref{dy:sec:Appendix_MeanVariance}
details this evaluation when when the number of users $N_{s,\ell}$
of slice $s$ at time slot $\ell$ is described by a binomial distribution.
Appendix~\ref{dy:sec:Appendix_PSP} describes the choice of $\gamma_{s,\ell}$
such that the satisfaction of (\ref{dy:eq:Relax:Satisfy:Node}, \ref{dy:eq:Relax:Satisfy:Link})
implies that of~\eqref{dy:eq:Cplx:Cons:Success}. This way, the MNO
can control the bound of the probability of SLA non-satisfaction.
The quantities $\widehat{R}_{s,n,\ell}$ and $\widehat{R}_{s,\text{b},\ell}$
are transmitted by the MNO to the InP as part of the MI-SLA.

Similarly, the InP translates the IP constraints (\ref{dy:eq:Cplx:Cons:Impact:Node},
\ref{dy:eq:Cplx:Cons:Impact:Link}), $\forall\left(i,ij\right)\in\mathcal{G}$
and $\forall n\in\Upsilon$, into 
\begin{align}
 & \sum_{s\in\mathcal{R}_{k}\cup\mathcal{S}_{k\textrm{-}1}^{\text{p}},v\in\mathcal{N}_{s}}\hspace{-0.8cm}\kappa_{s,\ell}\left(i,v\right)r_{s,n}\left(v\right)\leqslant a_{n}\left(i\right)-\widehat{B}_{n,\ell}\left(i,\gamma_{\text{B},\ell}\right),\label{dy:eq:Relax:Limit:Node}\\
 & \sum_{s\in\mathcal{R}_{k}\cup\mathcal{S}_{k\textrm{-}1}^{\text{p}},vw\in\mathcal{E}_{s}}\hspace{-0.9cm}\kappa_{s,\ell}\left(ij,vw\right)r_{s,\text{b}}\left(vw\right)\leqslant a_{\text{b}}\left(ij\right)-\widehat{B}_{\text{b},\ell}\left(ij,\gamma_{\text{B},\ell}\right),\label{dy:eq:Relax:Limit:Link}
\end{align}
for $\ell>k$, where 
\begin{align}
\widehat{B}_{n,\ell}\left(i,\gamma_{\text{B},\ell}\right) & =\overline{B}_{n,\ell}\left(i\right)+\gamma_{\text{B},\ell}\widetilde{B}_{n,\ell}\left(i\right),\label{dy:eq:MILP:Bound:Bn}\\
\widehat{B}_{\text{b},\ell}\left(ij,\gamma_{\text{B},\ell}\right) & =\overline{B}_{\text{b},\ell}\left(ij\right)+\gamma_{\text{B},\ell}\widetilde{B}_{\text{b},\ell}\left(ij\right)\label{dy:eq:MILP:Bound:Bb}
\end{align}
are the considered target level of background service demands. The
parameter $\gamma_{\text{B},\ell}>0$ has to be chosen such that the
satisfaction of the constraints~(\ref{dy:eq:Relax:Limit:Node}, \ref{dy:eq:Relax:Limit:Link})
implies the satisfaction of the IP constraints (\ref{dy:eq:Cplx:Cons:Impact:Node},
\ref{dy:eq:Cplx:Cons:Impact:Link}), see Appendix~\ref{dy:sec:Appendix_IP}
for more details. Moreover, if the constraints~(\ref{dy:eq:Relax:Limit:Node},
\ref{dy:eq:Relax:Limit:Link}) are satisfied by some assignment $\boldsymbol{\kappa}_{\mathcal{R}_{k}}$,
then the conditions \eqref{dy:eq:Cplx:Cons:Limit:Node-bis} and \eqref{dy:eq:Cplx:Cons:Limit:Link-bis}
are also satisfied.

\subsection{Costs and Incomes\label{dy:subsec:Costs-and-incomes}}

Consider the processing time interval $\mathcal{P}_{k}$ during which
a resource reservation scheme for all slices $s\in\mathcal{R}_{k}$
has to be evaluated. This amounts at evaluating $\mathbf{d}_{\mathcal{R}_{k}}$
and the assignments $\boldsymbol{\kappa}_{\mathcal{R}_{k}}$.

The costs charged by the InP to the MNO for a reservation scheme for
slice $s\in\mathcal{R}_{k}$ described by $\boldsymbol{\kappa}_{s,\ell}$
in time slot $\ell$ are spread between node and bandwidth \emph{resource}
reservation costs 
\begin{align}
C_{\text{r}}\left(\boldsymbol{\kappa}_{s,\ell}\right) & =\sum_{i\in\mathcal{N}}\sum_{v\in\mathcal{N}_{s}}\sum_{n\in\Upsilon}\kappa_{s,\ell}\left(i,v\right)r_{n}\left(v\right)c_{n}\left(i\right)\nonumber \\
 & +\sum_{ij\in\mathcal{E}}\sum_{vw\in\mathcal{E}_{s}}\kappa_{s,\ell}\left(ij,vw\right)r_{\text{b}}\left(vw\right)c_{\text{b}}\left(ij\right)\label{dy:eq:TotalResCosts}
\end{align}
as well as \emph{fixed} node disposal costs
\begin{equation}
C_{\text{f}}\left(\boldsymbol{\kappa}_{s,\ell}\right)=\sum_{i\in\mathcal{N}}\widetilde{\kappa}_{s,\ell}\left(i\right)c_{\textrm{f}}\left(i\right)\label{dy:eq:NodeDispCosts}
\end{equation}
for the infrastructure nodes used, where
\begin{align}
\widetilde{\kappa}_{s,\ell}\left(i\right) & =\begin{cases}
1 & \text{if }\sum_{v\in\mathcal{N}_{s}}\kappa_{s,\ell}\left(i,v\right)>0\\
0 & \text{otherwise},
\end{cases}\label{dy:eq:KappaTilde}
\end{align}
indicates whether node $i$ is used by slice $s$ in time slot $\ell$.

Additionally, when the amount of reserved resources for slice $s$
increases during two consecutive time slots, resource variation adaptation
costs are also charged by the InP to the MNO
\begin{align}
 & C_{\text{a}}\left(\boldsymbol{\kappa}_{s,\ell},\boldsymbol{\kappa}_{s,\ell-1}\right)\nonumber \\
 & =\sum_{i\in\mathcal{N}}\sum_{v\in\mathcal{N}_{s}}\max\left\{ \kappa_{s,\ell}\left(i,v\right)-\kappa_{s,\ell\textrm{-}1}\left(i,v\right),0\right\} c_{\textrm{a}}\left(i\right),\label{dy:eq:PreparationCosts}
\end{align}
see Section~\ref{subsec:Slice-preparation}.

Once a reservation request for a slice $s\in\mathcal{R}_{k}$ has
been granted by the InP, the MNO will be able to deploy the slice
(see the \emph{commissioning} and \emph{operation} blocks of Figure~\ref{dy:fig:Intro:SliceLifeCycle})
and receives from the SP some income $I_{s}$ depending on the complexity
and of the load of the slice.\vspace{-0.2cm}

\subsection{Optimization Problem\label{dy:subsec:OptimProb}}

For a given assignment $\boldsymbol{\mathbf{\kappa}}_{\mathcal{R}_{k}}$,
the earnings of the InP are the costs charged to the MNOs
\begin{align}
E_{k}^{\text{InP}}\left(\boldsymbol{\kappa}_{\mathcal{R}_{k}}\right)=\sum_{s\in\mathcal{R}_{k}}\sum_{\ell\in\mathcal{K}_{s}} & \left(C_{\text{r}}\left(\boldsymbol{\kappa}_{s,\ell}\right)+C_{\text{f}}\left(\boldsymbol{\kappa}_{s,\ell}\right)\right.\nonumber \\
 & \left.+C_{\text{a}}\left(\boldsymbol{\kappa}_{s,\ell},\boldsymbol{\kappa}_{s,\ell-1}\right)\right).\label{dy:eq:E_INP}
\end{align}
The InP may be interested in an assignment $\boldsymbol{\kappa}_{\mathcal{R}_{k}}$
that maximizes $E_{k}^{\text{InP}}\left(\boldsymbol{\kappa}_{\mathcal{R}_{k}}\right)$.
Nevertheless, such assignment would be detrimental for the earnings
of the MNOs expressed as
\begin{align}
E_{k}^{\text{MNO}}\left(\boldsymbol{\kappa}_{\mathcal{R}_{k}}\right)= & \sum_{s\in\mathcal{R}_{k}}d_{s}I_{s}-E_{k}^{\text{InP}}\left(\boldsymbol{\kappa}_{\mathcal{R}_{k}}\right).\label{dy:eq:E_MNO}
\end{align}
Consequently, MNOs may not be interested by InPs applying an optimization
strategy trying to maximizing $E_{k}^{\text{InP}}\left(\boldsymbol{\kappa}_{\mathcal{R}_{k}}\right)$.

Alternatively, the InP may try to find an assignment which maximizes
$E_{k}^{\text{MNO}}\left(\boldsymbol{\kappa}_{\mathcal{R}_{k}}\right)$.
This approach reduces the per-slice income for the InP, but allows
more slice resource reservation requests to be granted. Nevertheless,
InPs are usually unaware of the income $I_{s}$ obtained by the MNOs
from the SP, therefore $E_{k}^{\text{MNO}}\left(\boldsymbol{\kappa}_{\mathcal{R}_{k}}\right)$
cannot be evaluated by the InP. 

Consequently, we will consider an approach where the InP tries, for
a given $\mathbf{d}_{\mathcal{R}_{k}}$ (which determines\emph{ }the
number of accepted slices), to find an assignment $\boldsymbol{\kappa}_{\mathcal{R}_{k}}$
minimizing the reservation costs charged to the MNO. Moreover, the
InP also tries to maximize, with respect to $\mathbf{d}_{\mathcal{R}_{k}}$,
its earning $\min_{\boldsymbol{\kappa}_{\mathcal{R}_{k}}}\,E_{k}^{\text{InP}}\left(\boldsymbol{\kappa}_{\mathcal{R}_{k}}\right)$.
This approach leads to a max-min optimization problem. Its solution
provides the InP maximum earnings for granting slice requests and
is also appropriate for the MNO in terms of charged reservation costs.
Moreover, this approach potentially saves infrastructure resources
to satisfy future slice resource reservation requests.

Consequently, the \emph{joint} reservation of resources for all slices~$s\in\mathcal{R}_{k}$
during the processing time interval $\mathcal{P}_{k}$ is formulated
as Problem~\ref{dy:Prob:1}.

\LinesNumberedHidden{\begin{alProblem}

\caption{Max-Min Joint Slice Resource Reservation\hspace*{-0.2cm}\label{dy:Prob:1}}\vspace{-0.2cm}

\begin{align*}
 & \hspace{-0.4cm}\underset{\mathbf{d}_{\mathcal{R}_{k}}}{\mathrm{max}}\:\underset{\boldsymbol{\kappa}_{\mathcal{R}_{k}}}{\mathrm{min}}\:E_{k}^{\text{InP}}\left(\boldsymbol{\kappa}_{\mathcal{R}_{k}}\right)\\
 & \hspace{-0.4cm}\mathrm{s.t.}\,\forall\ell>k:\\
 & \hspace{-0.2cm}\sum\limits _{i\in\mathcal{N}}\kappa_{s,\ell}\left(i,v\right)r_{s,n}\left(v\right)\geqslant d_{s}\widehat{R}_{s,n,\ell}\left(v,\gamma_{s,\ell}\right),\forall n\in\Upsilon,v\in\mathcal{N}_{s},\\
 & \hspace{-0.2cm}\sum\limits _{ij\in\mathcal{E}}\kappa_{s,\ell}\left(ij,vw\right)r_{s,\text{b}}\left(vw\right)\geqslant d_{s}\widehat{R}_{s,\text{b},\ell}\left(vw,\gamma_{s,\ell}\right),\forall vw\in\mathcal{E}_{s},\\
 & \hspace{-0.37cm}\text{and s.t. }\forall\ell>k,\forall i\in\mathcal{N}:\\
 & \hspace{-0.2cm}\sum_{s\in\mathcal{R}_{k}\cup\mathcal{S}_{k\textrm{-}1}^{\text{p}},v\in\mathcal{N}_{s}}\hspace{-0.8cm}\kappa_{s,\ell}\left(i,v\right)r_{s,n}\left(v\right)\leqslant a_{n}\left(i\right)-\widehat{B}_{n,\ell}\left(i,\gamma_{\text{B},\ell}\right),\\
 & \hspace{-0.37cm}\text{and s.t. }\forall\ell>k,\forall ij\in\mathcal{E}:\\
 & \hspace{-0.2cm}\sum_{s\in\mathcal{R}_{k}\cup\mathcal{S}_{k\textrm{-}1}^{\text{p}},vw\in\mathcal{E}_{s}}\hspace{-0.9cm}\kappa_{s,\ell}\left(ij,vw\right)r_{s,\text{b}}\left(vw\right)\leqslant a_{\text{b}}\left(ij\right)-\widehat{B}_{\text{b},\ell}\left(ij,\gamma_{\text{B},\ell}\right),\\
 & \hspace{-0.2cm}\sum\limits _{j\in\mathcal{N}}\left[\kappa_{s,\ell}\left(ij,vw\right)-\kappa_{s,\ell}\left(ji,vw\right)\right]=\\
 & \hspace{-0.4cm}\Bigg(\frac{r_{s,\text{b}}(vw)}{\sum\limits _{vu\in\mathcal{E}_{s}}r_{s,\text{b}}(vu)}\Bigg)\kappa_{s,\ell}\left(i,v\right)-\Bigg(\frac{r_{s,\text{b}}(vw)}{\sum\limits _{uw\in\mathcal{E}_{s}}r_{s,\text{b}}(uw)}\Bigg)\kappa_{s,\ell}\left(i,w\right),\\
 & \hspace{-0.2cm}\forall s\in\mathcal{R}_{k},i\in\mathcal{N},\forall vw\in\mathcal{E}_{s}.
\end{align*}
\end{alProblem}}

The max-min optimization makes Problem~\ref{dy:Prob:1} difficult
to solve. This motivates us to develop some efficient heuristics in
Section~\ref{dy:sec:ProvAlgo}.\vspace{-0.2cm}

\section{Slice Resource Reservation Algorithms\label{dy:sec:ProvAlgo}}

In this section, two heuristics are introduced to provide approximate
solutions to Problem~\ref{dy:Prob:1}, performing either joint or
sequential slice resource reservation.

\subsection{Linearization of the Cost Function\label{dy:subsec:Relax:MaxMin}}

In \eqref{dy:eq:E_INP}, the term $C_{\text{a}}\left(\boldsymbol{\kappa}_{s,\ell},\boldsymbol{\kappa}_{s,\ell-1}\right)$
makes the objective function nonlinear. To address this issue, consider
the set of variables
\begin{equation}
\mathbf{y}_{s}=\left\{ y_{s,\ell}\left(i,v\right):\ell\in\mathcal{K}_{s},i\in\mathcal{N},v\in\mathcal{N}_{s}\right\} 
\end{equation}
for each $s\in\mathcal{R}_{k}$, $\mathbf{y}_{\mathcal{R}_{k}}=\left\{ \mathbf{y}_{s}:s\in\mathcal{R}_{k}\right\} $,
and reformulate the objective function \eqref{dy:eq:E_INP} as
\begin{align}
E_{k}^{\text{InP}}\left(\boldsymbol{\kappa}_{\mathcal{R}_{k}},\mathbf{y}_{\mathcal{R}_{k}}\right) & =\sum_{s\in\mathcal{R}_{k}}\sum_{\ell\in\mathcal{K}_{s}}\Big(C_{\text{r}}\left(\boldsymbol{\kappa}_{s,\ell}\right)+C_{\text{f}}\left(\boldsymbol{\kappa}_{s,\ell}\right)\nonumber \\
 & +\sum\limits _{i\in\mathcal{N}}\sum_{v\in\mathcal{N}_{s}}y_{s,\ell}\left(i,v\right)c_{\textrm{a}}\left(i\right)\Big)
\end{align}
with the additional constraints, to be satisfied for all $s\in\mathcal{R}_{k}$,
$\ell\in\mathcal{K}_{s}$, $i\in\mathcal{N}$, and $v\in\mathcal{N}_{s}$
\begin{align}
y_{s,\ell}\left(i,v\right) & \geqslant\kappa_{s,\ell}\left(i,v\right)-\kappa_{s,\ell\textrm{-}1}\left(i,v\right),\label{dy:eq:Minimax:Linearized:C1}\\
y_{s,\ell}\left(i,v\right) & \geqslant0.\label{dy:eq:Minimax:Linearized:C2}
\end{align}
For a given value of $\mathbf{d}_{\mathcal{R}_{k}}$, the objective
function has now to be minimized with respect to $\boldsymbol{\kappa}_{s,\ell}$,
$s\in\mathcal{R}_{k}$, $\ell\in\mathcal{K}_{s}$, and $\mathbf{y}_{\mathcal{R}_{k}}$.

Moreover, the evaluation of $C_{\text{f}}\left(\boldsymbol{\kappa}_{s,\ell}\right)$
involves $\widetilde{\kappa}_{s,\ell}\left(i\right)$ defined in \eqref{dy:eq:KappaTilde}.
The variable $\widetilde{\kappa}_{s,\ell}\left(i\right)$ can be related
to $\sum\limits _{v}\kappa_{s,\ell}\left(i,v\right)$ using the following
linear inequality constraints 
\begin{align}
\sum\limits _{v}\kappa_{s,\ell}\left(i,v\right) & \geqslant0,\\
\widetilde{\kappa}_{s,\ell}\left(i\right)\overline{N} & \geqslant\sum\limits _{v}\kappa_{s,\ell}\left(i,v\right),
\end{align}
where $\overline{N}$ is an upper bound on the number of VNF instances
of all types for which resources may be reserved by a given infrastructure
node.

\subsection{Relaxed Joint Max-Min Optimization Problem\label{dy:subsec:Relax:Joint}}

Even with the results of Section~\ref{dy:subsec:Relax:MaxMin}, the
solution of Problem~\ref{dy:Prob:1} requires addressing a constrained
max-min optimization problem, which is still quite complex. To address
this issue, for a fixed value of $\mathbf{d}_{\mathcal{R}_{k}}$,
we introduce the following optimization problem.

\LinesNumberedHidden{\begin{alProblem}

\caption{Joint Slice Resource Reservation Given $\mathbf{d}_{\mathcal{R}_{k}}$\hspace*{-0.8cm}\label{dy:Prob:2}}\vspace{-0.2cm}

\begin{align*}
 & \hspace{-0.45cm}\underset{\boldsymbol{\kappa}_{\mathcal{R}_{k}},\mathbf{y}_{\mathcal{R}_{k}}}{\mathrm{min}}\,E_{k}^{\text{InP}}\left(\boldsymbol{\kappa}_{\mathcal{R}_{k}},\mathbf{y}_{\mathcal{R}_{k}}\right)\\
 & \hspace{-0.37cm}\text{s.t.}\,\forall s\in\mathcal{R}_{k},\forall\ell>k:\\
 & \hspace{-0.2cm}\sum\limits _{i\in\mathcal{N}}\kappa_{s,\ell}\left(i,v\right)r_{s,n}\left(v\right)\geqslant d_{s}\widehat{R}_{s,n,\ell}\left(v,\gamma_{s,\ell}\right),\forall n\in\Upsilon,v\in\mathcal{N}_{s},\\
 & \hspace{-0.2cm}\sum\limits _{ij\in\mathcal{E}}\kappa_{s,\ell}\left(ij,vw\right)r_{s,\text{b}}\left(vw\right)\geqslant d_{s}\widehat{R}_{s,\text{b},\ell}\left(vw,\gamma_{s,\ell}\right),\forall vw\in\mathcal{E}_{s},\\
 & \hspace{-0.1cm}y_{s,\ell}\left(i,v\right)\geqslant\kappa_{s,\ell}\left(i,v\right)-\kappa_{s,\ell\textrm{-}1}\left(i,v\right),\forall i\in\mathcal{N},v\in\mathcal{N}_{s},\\
 & \hspace{-0.1cm}y_{s,\ell}\left(i,v\right)\geqslant0,\forall i\in\mathcal{N},v\in\mathcal{N}_{s},\\
 & \hspace{-0.37cm}\text{and s.t. }\forall\ell>k,\forall i\in\mathcal{N}:\\
 & \hspace{-0.2cm}\sum_{s\in\mathcal{R}_{k}\cup\mathcal{S}_{k\textrm{-}1}^{\text{p}},v\in\mathcal{N}_{s}}\hspace{-0.8cm}\kappa_{s,\ell}\left(i,v\right)r_{s,n}\left(v\right)\leqslant a_{n}\left(i\right)-\widehat{B}_{n,\ell}\left(i,\gamma_{\text{B},\ell}\right),\\
 & \hspace{-0.37cm}\text{and s.t. }\forall\ell>k,\forall ij\in\mathcal{E}:\\
 & \hspace{-0.2cm}\sum_{s\in\mathcal{R}_{k}\cup\mathcal{S}_{k\textrm{-}1}^{\text{p}},vw\in\mathcal{E}_{s}}\hspace{-0.9cm}\kappa_{s,\ell}\left(ij,vw\right)r_{s,\text{b}}\left(vw\right)\leqslant a_{\text{b}}\left(ij\right)-\widehat{B}_{\text{b},\ell}\left(ij,\gamma_{\text{B},\ell}\right),\\
 & \hspace{-0.2cm}\sum\limits _{j\in\mathcal{N}}\left[\kappa_{s,\ell}\left(ij,vw\right)-\kappa_{s,\ell}\left(ji,vw\right)\right]=\\
 & \hspace{-0.4cm}\Bigg(\frac{r_{s,\text{b}}(vw)}{\sum\limits _{vu\in\mathcal{E}_{s}}r_{s,\text{b}}(vu)}\Bigg)\kappa_{s,\ell}\left(i,v\right)-\Bigg(\frac{r_{s,\text{b}}(vw)}{\sum\limits _{uw\in\mathcal{E}_{s}}r_{s,\text{b}}(uw)}\Bigg)\kappa_{s,\ell}\left(i,w\right),\\
 & \hspace{-0.2cm}\forall s\in\mathcal{R}_{k},i\in\mathcal{N},\forall vw\in\mathcal{E}_{s}.
\end{align*}
\end{alProblem}}

One considers then a greedy solution approach to Problem~\ref{dy:Prob:1},
where the slices to be processed in $\mathcal{R}_{k}=\left\{ s_{1},\dots,s_{R_{k}}\right\} $
are assumed to be ordered, see Section~\ref{dy:subsec:Provisioning-Scheduling}.
When several MNOs have submitted slice resource reservation requests
in the same time slot, the InP may also prioritize MNOs.

\subsection{Relaxed Single Slice Max-Min Optimization Problem\label{dy:subsec:Relax:Single}}

The number of variables involved in Problem~\ref{dy:Prob:2} introduced
in Section~\ref{dy:subsec:Relax:Joint} may become relatively large
when the resource reservation is performed simultaneously for several
slices. For this reason, we introduce a reduced-complexity version
of Problem~\ref{dy:Prob:2}, where the resource reservation is performed
slice by slice. One focuses on a slice~$s\in\mathcal{R}_{k}$ which
resource reservation request has to be processed. Some resource reservation
requests for slices $s'\in\mathcal{R}_{k}$, $s'\neq s$ may have
been previously processed, in which case, when the request is granted,
$d_{s'}=1$ and $\boldsymbol{\kappa}_{s'}\neq\mathbf{0}$ and when
it is not granted, $d_{s'}=0$ and $\boldsymbol{\kappa}_{s'}=\mathbf{0}$.
For not yet processed requests of slices $s'\in\mathcal{R}_{k}$,
$s'\neq s$, one considers $d_{s'}=0$ and $\boldsymbol{\kappa}_{s'}=\mathbf{0}$.
With these assumptions, reserving resources for slice $s\in\mathcal{R}_{k}$
requires the solution of Problem~\ref{dy:Prob:3}.

\LinesNumberedHidden{\begin{alProblem}

\caption{Single Slice Resource Reservation\label{dy:Prob:3}}\vspace{-0.2cm}

\begin{align*}
 & \hspace{-0.45cm}\max_{d_{s}}\min_{\boldsymbol{\kappa}_{s},\mathbf{y}_{s}}\,E_{k}^{\text{InP}}\left(\boldsymbol{\kappa}_{s},\mathbf{y}_{s}\right)\\
 & \hspace{-0.37cm}\textrm{s.t. }\forall\ell\in\mathcal{K}_{s}:\\
 & \hspace{-0.2cm}\sum\limits _{i\in\mathcal{N}}\kappa_{s,\ell}\left(i,v\right)r_{s,n}\left(v\right)\geqslant d_{s}\widehat{R}_{s,n,\ell}\left(v,\gamma_{s,\ell}\right),\forall n\in\Upsilon,v\in\mathcal{N}_{s},\\
 & \hspace{-0.2cm}\sum\limits _{ij\in\mathcal{E}}\kappa_{s,\ell}\left(ij,vw\right)r_{s,\text{b}}\left(vw\right)\geqslant d_{s}\widehat{R}_{s,\text{b},\ell}\left(vw,\gamma_{s,\ell}\right),\forall vw\in\mathcal{E}_{s},\\
 & \hspace{-0.1cm}y_{s,\ell}\left(i,v\right)\geqslant\kappa_{s,\ell}\left(i,v\right)-\kappa_{s,\ell\textrm{-}1}\left(i,v\right),\forall i\in\mathcal{N},v\in\mathcal{N}_{s},\\
 & \hspace{-0.1cm}y_{s,\ell}\left(i,v\right)\geqslant0,\forall i\in\mathcal{N},v\in\mathcal{N}_{s},\\
 & \hspace{-0.37cm}\text{and s.t. }\forall\ell\in\mathcal{K}_{s},\forall i\in\mathcal{N}:\\
 & \hspace{-0.2cm}\sum_{s'\in\mathcal{R}_{k}\cup\mathcal{S}_{k\textrm{-}1}^{\text{p}},v\in\mathcal{N}_{s}}\hspace{-0.8cm}\kappa_{s'\hspace{-0.06cm},\ell}\left(i,v\right)r_{s'\hspace{-0.06cm},n}\left(v\right)\leqslant a_{n}\left(i\right)-\widehat{B}_{n,\ell}\left(i,\gamma_{\text{B},\ell}\right),\\
 & \hspace{-0.37cm}\text{and s.t. }\forall\ell\in\mathcal{K}_{s},\forall ij\in\mathcal{E}:\\
 & \hspace{-0.3cm}\sum_{s'\in\mathcal{R}_{k}\cup\mathcal{S}_{k\textrm{-}1}^{\text{p}},vw\in\mathcal{E}_{s}}\hspace{-0.9cm}\kappa_{s'\hspace{-0.06cm},\ell}\left(ij,vw\right)r_{s'\hspace{-0.06cm},\text{b}}\left(vw\right)\leqslant a_{\text{b}}\left(ij\right)-\widehat{B}_{\text{b},\ell}\left(ij,\gamma_{\text{B},\ell}\right),\\
 & \hspace{-0.2cm}\sum\limits _{j\in\mathcal{N}}\left[\kappa_{s,\ell}\left(ij,vw\right)-\kappa_{s,\ell}\left(ji,vw\right)\right]=\\
 & \hspace{-0.4cm}\Bigg(\frac{r_{s,\text{b}}(vw)}{\sum\limits _{vu\in\mathcal{E}_{s}}r_{s,\text{b}}(vu)}\Bigg)\kappa_{s,\ell}\left(i,v\right)-\Bigg(\frac{r_{s,\text{b}}(vw)}{\sum\limits _{uw\in\mathcal{E}_{s}}r_{s,\text{b}}(uw)}\Bigg)\kappa_{s,\ell}\left(i,w\right),\\
 & \hspace{-0.2cm}\forall i\in\mathcal{N},vw\in\mathcal{E}_{s}.
\end{align*}
\end{alProblem}}

Assuming again that the slice resource reservation requests are ordered,
see Section~\ref{dy:subsec:Provisioning-Scheduling}, one may get
a second greedy reservation algorithm where slice resource reservation
requests are processed slice by slice solving Problem~\ref{dy:Prob:3}
for each slice. The highest priority slice is processed first. The
lower priority slices are then processed, whatever the reservation
result of a higher priority slice. Even if high-priority slices may
have their resource reservation request rejected, lower-priority slice
requests may be granted for slices with smaller resource requirements.

\subsection{Slice Resource Reservation Approaches \label{dy:subsec:Prov:Approaches}}

For the suboptimal algorithms introduced in Sections~\ref{dy:subsec:Relax:Joint}
and \ref{dy:subsec:Relax:Single}, two Prioritized slice resource
Reservation ($\mathtt{PR}$) variants are considered, depending on
whether slices resource reservation requests are processed jointly
($\mathtt{J\text{-}PR}$) or sequentially ($\mathtt{S\text{-}PR}$).

\subsubsection{Joint Approach}

In the $\mathtt{J\text{-}PR}$ approach, all slices in $\mathcal{R}_{k}$
are processed jointly. This is done by solving Problem~\ref{dy:Prob:2},
starting by fixing $\mathbf{d}_{\mathcal{R}_{k}}=\left(d_{1},\dots,d_{R_{k}}\right)=\left(1,\dots,1\right)$,
\emph{i.e.}, we try initially to satisfy all slice resource reservation
requests. If the reservation is successful, the algorithm stops. If
no solution is returned, the resource reservation request of the slice
with lowest priority is not granted, \emph{i.e.}, $d_{R_{k}}=0$.
Problem~\ref{dy:Prob:2} is solved again considering $\mathbf{d}_{\mathcal{R}_{k}}=\left(1,\dots,1,0\right)$.
If there is still no solution, the resource reservation request for
the slice with second lowest priority is not granted, and so forth.
If more than two slice requests have the same lowest priority, the
last arrived one is not granted. The first part of Algorithm~\ref{dy:algo:All}
(Lines~4\textendash 13) summarizes the $\mathtt{J\text{-}PR}$ approach.

\subsubsection{Sequential Approach}

In the $\mathtt{S\text{-}PR}$ approach, slice resource reservation
requests in $\mathcal{R}_{k}$ are processed sequentially. This is
done by solving Problem~\ref{dy:Prob:3}, for each slice $s\in\mathcal{R}_{k}$,
$d_{s}\in\left\{ 0,1\right\} $, starting from that with highest priority.
The maximization of the cost function considers $d_{s}=1$. If the
following minimization problem admits a solution, one keeps $d_{s}=1$.
If the minimization problem admits no solution, $d_{s}=0$ and $\boldsymbol{\kappa}_{s}=\mathbf{0}$
is the solution.

The second part of Algorithm~\ref{dy:algo:All} (Lines~14\textendash 17)
summarizes the $\mathtt{S\text{-}PR}$ approach. Note that, $\mathtt{S\text{-}PR}$
when $\alpha=0$ implements a first-arrived first-served processing
policy.

\begin{algorithm} \footnotesize

\caption{ Prioritized Slice Resource Reservation$\hspace{-0.5cm}$\label{dy:algo:All}}

\BlankLine

\everypar={\nl}

\ForEach{ processing time interval $\mathcal{P}_{k}$ }{

Get sorted slice request set $\mathcal{R}_{k}$ from MNO;

\Switch{ $\mathtt{reservation\_variant}$}{

\Case{ $\mathtt{J\text{-}PR}$ (joint prioritized reservation) }{

Initialize $\mathbf{d}_{\mathcal{R}_{k}}=\left(1,\dots,1\right)$;

$i=\left|\mathcal{R}_{k}\right|$;

\While{ $i>0$}{

Solve Problem~\ref{dy:Prob:2} to get $\boldsymbol{\kappa}_{\mathcal{R}_{k}}$;

\eIf{no feasible solution found}{

$d_{i}=0$;

$i=i-1;$

}{

break;

} 

} 

} 

\Case{ $\mathtt{S\text{-}PR}$ (sequential prioritized reservation)
}{

Initialize $d_{s}=0$, $\boldsymbol{\kappa}_{s}=\mathbf{0}$, $\forall s\in\mathcal{R}_{k}$;

\ForEach{$s\in\mathcal{R}_{k}$}{

Solve Problem~\ref{dy:Prob:3} for slice~$s$ to get $d_{s}$ and
$\boldsymbol{\kappa}_{s}$;

} 

} 

} 

\textit{\# Update flag of processed slice requests}

Set $f_{s}=1$, $\forall s\in\mathcal{R}_{k}$;

} 

\end{algorithm}

\subsubsection{Complexity Analysis\label{subsec:Complexity-Analysis}}

Each variant in Algorithm~\ref{dy:algo:All} requires the solution
of one or several ILPs, whose complexity increases exponentially with
the number of variables in the worst case. The $\mathtt{J\text{-}PR}$
variant considers a single ILP involving $\sum_{s\in\mathcal{R}_{k}}\sum_{\ell\in\mathcal{K}_{s}}\left(\left|\mathcal{N}\right|+2\left|\mathcal{N}\right|\left|\mathcal{N}_{s}\right|+\left|\mathcal{E}\right|\left|\mathcal{E}_{s}\right|\right)$
variables and $\sum_{s\in\mathcal{R}_{k}}\sum_{\ell\in\mathcal{K}_{s}}\left(\left|\mathcal{N}\right|\left|\mathcal{E}_{s}\right|+\left|\mathcal{N}_{s}\right|\left|\Upsilon\right|+\left|\mathcal{E}_{s}\right|+2\left|\mathcal{N}\right|\left|\mathcal{N}_{s}\right|\right)+\sum_{\ell\in\mathcal{K}_{s}}\left(\left|\mathcal{N}\right|+\left|\mathcal{E}\right|\right)$
constraints. The $\mathtt{S\text{-}PR}$ variant instead splits the
task into $\left|\mathcal{R}_{k}\right|$ subproblems, each of which
involves $\sum_{\ell\in\mathcal{K}_{s}}\left(\left|\mathcal{N}\right|+2\left|\mathcal{N}\right|\left|\mathcal{N}_{s}\right|+\left|\mathcal{E}\right|\left|\mathcal{E}_{s}\right|\right)$
variables and $\sum_{\ell\in\mathcal{K}_{s}}\left(\left|\mathcal{N}\right|\left|\mathcal{E}_{s}\right|+\left|\mathcal{N}_{s}\right|\left|\Upsilon\right|+\left|\mathcal{E}_{s}\right|+2\left|\mathcal{N}\right|\left|\mathcal{N}_{s}\right|+\left|\mathcal{N}\right|+\left|\mathcal{E}\right|\right)$
constraints. Therefore, each subproblem in the sequential approach
implies $\left|\mathcal{R}_{k}\right|$ times less variables than
the joint variant. Consequently, due to the exponential complexity
of the NP-hard ILP, the sequential approach may provide a solution
faster than the joint variant. In Section~\ref{dy:sec:Evaluation},
the two proposed variants are compared via numerical simulations.

\section{Evaluation\label{dy:sec:Evaluation}}

This section presents simulations to evaluate the performance of the
two reservation algorithms, $\mathtt{J\text{-}PR}$ and $\mathtt{S\text{-}PR}$,
described in Section~\ref{dy:sec:ProvAlgo}. The simulation setup
is described in Section~\ref{dy:subsec:SetUp}. All simulations described
in Section~\ref{dy:subsec:Eva:Results} are performed with the CPLEX
MILP solver interfaced with MATLAB. Our work focuses on the slice
admission control and resource reservation mechanisms, both taking
place before any slice deployment and activation. Consequently, in
the simulation, aspects related to user admission control, radio coverage/interference,
or packet queuing and propagation delays are not considered.

\subsection{Simulation Conditions\label{dy:subsec:SetUp}}

\subsubsection{Infrastructure Topology}

\begin{figure}[tbh]
\begin{centering}
\includegraphics[width=0.6\columnwidth]{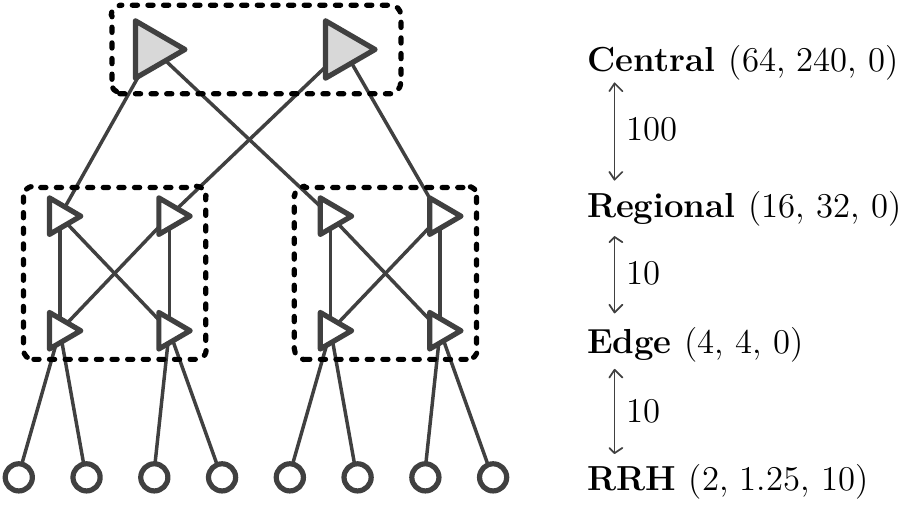}
\par\end{centering}
\caption{Description of the binary fat-tree infrastructure network considered
in the simulations; Nodes provide a given amount of computing $a_{\textrm{c}}$,
memory $a_{\textrm{m}}$, and wireless $a_{\textrm{w}}$ resources
expressed in number of used CPUs, Gbytes, and Gbps; Links are able
to transmit data at a rate $a_{\text{b}}$ expressed in Gbps.\label{dy:fig:Setup:Infra}}
\end{figure}

The considered infrastructure network is represented by the binary
fat tree topology depicted in Figure~\ref{dy:fig:Setup:Infra}, taken
from \cite{Riggio2016,Bouten2017}. The leaf nodes represent the far-edge
hosts of Radio Unit (RU)/Distributed Unit (DU). The other nodes represent
the edge, regional, and central data centers. Infrastructure nodes
provide a given amount of computing, storage, and possibly wireless
resources $\left(a_{\textrm{c}},a_{\textrm{m}},a_{\textrm{w}}\right)$,
expressed in number of CPUs, Gbytes, and Gbps, depending on the layer
they are located. Infrastructure links provide a given bandwidth $a_{\text{b}}$,
expressed in Gbps. The per-unit resource cost ($c_{n}\left(i\right)$
and $c_{\text{b}}\left(ij\right)$), fixed cost $c_{\text{f}}\left(i\right)$,
and reservation adaptation cost $c_{\text{a}}\left(i\right)$ are
respectively $1$, $10$, and $20$, $\forall\left(i,ij\right)\in\mathcal{G}$.

\subsubsection{Slice Resource Demand (S-RD)\label{dy:subsec:Eva:SRD}}

\begin{figure}[tbh]
\begin{centering}
\subfloat[\label{dy:fig:Eva:Setup:Pattern1}]{\begin{centering}
\includegraphics[width=0.47\columnwidth]{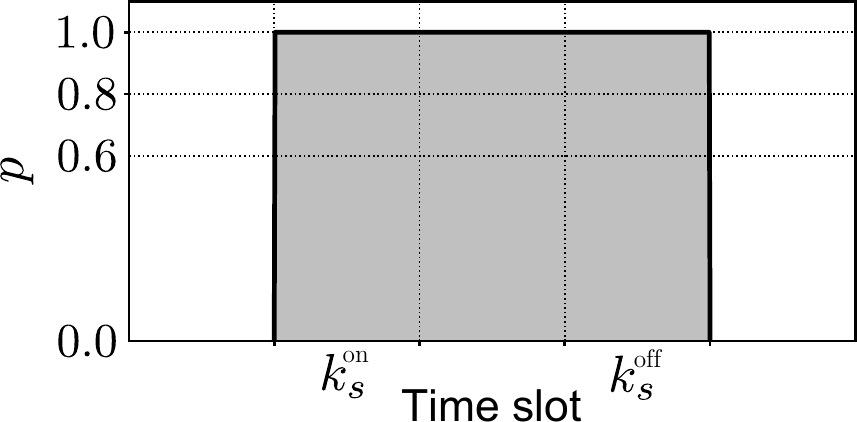}
\par\end{centering}
}\subfloat[\label{dy:fig:Eva:Setup:Pattern2}]{\begin{centering}
\includegraphics[width=0.47\columnwidth]{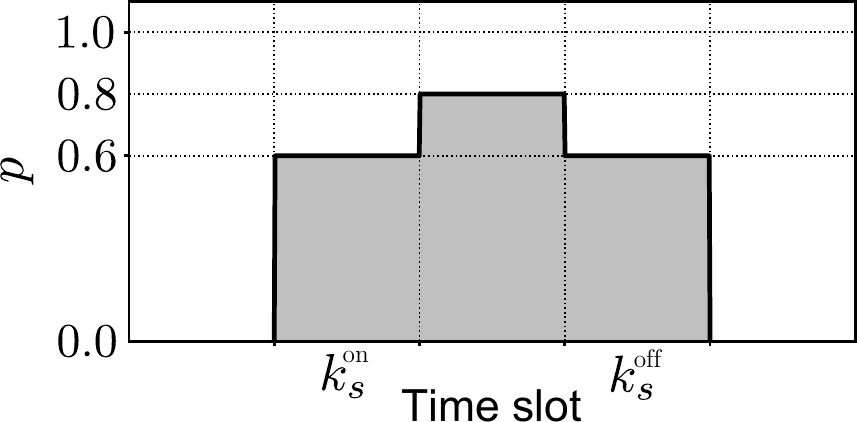}
\par\end{centering}
}
\par\end{centering}
\caption{Probability pattern of service usage: (a) constant over a time interval
and (b) piece-wise constant.\label{dy:fig:Eva:Setup:Pattern}}
\end{figure}

The number of users of a slice $s$ is assumed to follow a binomial
distribution of parameter $p_{s,k}$, see \cite{Dong2016,George2019}.
Considering a Gaussian distribution for the individual user resource
demands, the resulting resource demand for the slice follows a distribution
close to a log-normal distribution, as observed in \cite{Zhao2017}.
One considers two patterns to represent the evolution with time of
$p_{s,k}$, which impacts the evolution of the slice resource demands.
The first, illustrated in Figure~\ref{dy:fig:Eva:Setup:Pattern1},
corresponds to a constant demand $p_{s,k}=1$ during the whole lifetime
of the slice. The second, shown in Figure~\ref{dy:fig:Eva:Setup:Pattern2},
describes a slice whose resource demand evolves from one time slot
to the next.

Three types of slices are considered.
\begin{itemize}
\item Slices of type~1 aim to provide an HD video streaming service at
an average rate of $6$~Mbps for VIP users, \emph{e.g.}, in a stadium.
The number of users of a slice $s$ of type~$1$ follows a binomial
distribution $\mathcal{B}\left(500,p_{s,k}\right)$. The function
architecture of slices of type~1 is composed of $3$ VNFs: a virtual
Video Optimization Controller (vVOC), a virtual Gateway (vGW), and
a virtual Base Band Unit (vBBU). The required SSP for type 1-slices
is $\underline{p}_{s}^{\text{sp}}=0.99$;
\item Slices of type~2 are dedicated to provide an SD video streaming service
at an average rate of $4$~Mbps. The number of users of a slice $s$
of type~$2$ follows a binomial distribution $\mathcal{B}\left(2000,p_{s,k}\right)$.
The function architecture of slices of type~2 is similar to that
of type~1. The required SSP for type 2-slices is $\underline{p}_{s}^{\text{sp}}=0.95$;
\item Slices of type~3 aim to provide a video surveillance and traffic
monitoring service at an average rate of $2$~Mbps for $200$ cameras,
\emph{e.g.}, installed along a highway. The demand pattern of such
type of slice is thus always that of Figure~\ref{dy:fig:Eva:Setup:Pattern1}.
The third slice type consists of five virtual functions: a vBBU, a
vGW, a virtual Traffic Monitor (vTM), a vVOC, and a virtual Intrusion
Detection Prevention System (vIDPS). The required SSP for type 3-slices
is $\underline{p}_{s}^{\text{sp}}=0.9$.
\end{itemize}
The slice type is chosen uniformly at random. For slices of type~$1$
and~$2$, the demand pattern is also chosen uniformly at random.

A normalized unit duration time slot is considered, \emph{i.e.}, $T=1$.
The processing duration is chosen as $\varepsilon T=0.1T$. The number
of reservation request arrivals in each time slot obeys a Poisson
distribution $\textrm{Pois}\left(\mu\right)$ of parameter $\mu=2$.
The arrival time of each slice request is uniformly distributed within
each time interval $\mathcal{T}_{k}$. The activation delay (\textit{i.e}.,
$k_{s}^{\text{on}}-k_{s}$) follows the uniform distribution $\mathcal{U}\left(1,6\right)$
and the lifetime follows the uniform distribution $\mathcal{U}\left(1,3\right)$.

As detailed in Section~\ref{dy:subsec:Slice-Resource-Demand}, the
resource requirements for the various SFCs that will have to be deployed
within a slice are aggregated within an S-RD graph that mimics the
SFC-RD graph. S-RD nodes and links are characterized by the aggregated
resources needed to support the targeted number of users. Details
of each resource type as well as the associated U-RD, SFC-RD, and
S-RD graph are given in Table~\ref{tab:SRD}. Numerical values in
Table~\ref{tab:SRD} have been adapted from \cite{Savi2017}. Each
slice request is randomly assigned one type among these three types.
\begin{center}
\begin{table*}[tbh]
\caption{Parameters of U-RD, SFC-RD, and S-RD \label{tab:SRD}}

\centering{}%
\begin{tabular}{lccccccc}
\toprule 
\multicolumn{8}{l}{\textbf{Type }1: HD video streaming at $6$~Mbps, $\underline{p}_{s}^{\text{sp}}=0.99$}\tabularnewline
\addlinespace[0.1cm]
\textit{Node} & $(\overline{U}_{s,\text{c}},\widetilde{U}_{s,\text{c}})$ & $(\overline{U}_{s,\text{m}},\widetilde{U}_{s,\text{m}})$ & $(\overline{U}_{s,\text{w}},\widetilde{U}_{s,\text{w}})$ & $(r_{\text{c}},r_{\text{m}},r_{\text{w}})$ & \textit{Link} & $(\overline{U}_{s,\text{b}},\widetilde{U}_{s,\text{b}})$ & $r_{s,\text{b}}$\tabularnewline
\cmidrule[0.4pt](lr{0.12em}){1-1}%
\cmidrule[0.4pt](lr{0.12em}){2-2}%
\cmidrule[0.4pt](lr{0.12em}){3-3}%
\cmidrule[0.4pt](lr{0.12em}){4-4}%
\cmidrule[0.4pt](lr{0.12em}){5-5}%
\cmidrule[0.4pt](lr{0.12em}){6-6}%
\cmidrule[0.4pt](lr{0.12em}){7-7}%
\cmidrule[0.4pt](lr{0.12em}){8-8}%
vVOC & $\left(5.4,0.54\right)\textrm{e-}3$ & $\left(1.5,0.15\right)\textrm{e-}2$ & \textemdash{} & $\left(0.29,0.81,0\right)$ & vVOC$\rightarrow$vGW & $\left(4,0.4\right)\textrm{e-}3$ & $0.22$\tabularnewline
vGW & $\left(9.0,0.90\right)\textrm{e-}4$ & $\left(5.0,0.50\right)\textrm{e-}4$ & \textemdash{} & $\left(0.05,0.03,0\right)$ & vGW$\rightarrow$vBBU & $\left(4,0.4\right)\textrm{e-}3$ & $0.22$\tabularnewline
vBBU & $\left(8.0,0.80\right)\textrm{e-}4$ & $\left(5.0,0.50\right)\textrm{e-}4$ & $\left(4,0.4\right)\textrm{e-}3$ & $\left(0.04,0.03,0.2\right)$ &  &  & \tabularnewline
\addlinespace[0.2cm]
\multicolumn{8}{l}{\textbf{Type 2}: SD video streaming at $4$~Mbps, $\underline{p}_{s}^{\text{sp}}=0.95$}\tabularnewline
\addlinespace[0.1cm]
\textit{Node} & $(\overline{U}_{s,\text{c}},\widetilde{U}_{s,\text{c}})$ & $(\overline{U}_{s,\text{m}},\widetilde{U}_{s,\text{m}})$ & $(\overline{U}_{s,\text{w}},\widetilde{U}_{s,\text{w}})$ & $(r_{\text{c}},r_{\text{m}},r_{\text{w}})$ & \textit{Link} & $(\overline{U}_{s,\text{b}},\widetilde{U}_{s,\text{b}})$ & $r_{s,\text{b}}$\tabularnewline
\cmidrule[0.4pt](lr{0.12em}){1-1}%
\cmidrule[0.4pt](lr{0.12em}){2-2}%
\cmidrule[0.4pt](lr{0.12em}){3-3}%
\cmidrule[0.4pt](lr{0.12em}){4-4}%
\cmidrule[0.4pt](lr{0.12em}){5-5}%
\cmidrule[0.4pt](lr{0.12em}){6-6}%
\cmidrule[0.4pt](lr{0.12em}){7-7}%
\cmidrule[0.4pt](lr{0.12em}){8-8}%
vVOC & $\left(1.1,0.11\right)\textrm{e-}3$ & $\left(7.5,0.75\right)\textrm{e-}3$ & \textemdash{} & $\left(0.17,1.20,0\right)$ & vVOC$\rightarrow$vGW & $\left(2,0.2\right)\textrm{e-}3$ & $0.32$\tabularnewline
vGW & $\left(1.8,0.18\right)\textrm{e-}4$ & $\left(2.5,0.25\right)\textrm{e-}4$ & \textemdash{} & $\left(0.03,0.04,0\right)$ & vGW$\rightarrow$vBBU & $\left(2,0.2\right)\textrm{e-}3$ & $0.32$\tabularnewline
vBBU & $\left(0.8,0.08\right)\textrm{e-}4$ & $\left(2.5,0.25\right)\textrm{e-}4$ & $\left(2,0.2\right)\textrm{e-}3$ & $\left(0.01,0.04,0.3\right)$ &  &  & \tabularnewline
\addlinespace[0.2cm]
\multicolumn{8}{l}{\textbf{Type 3}: Video surveillance and traffic monitoring at $2$~Mbps,
$\underline{p}_{s}^{\text{sp}}=0.9$}\tabularnewline
\addlinespace[0.1cm]
\textit{Node} & $(\overline{U}_{s,\text{c}},\widetilde{U}_{s,\text{c}})$ & $(\overline{U}_{s,\text{m}},\widetilde{U}_{s,\text{m}})$ & $(\overline{U}_{s,\text{w}},\widetilde{U}_{s,\text{w}})$ & $(r_{\text{c}},r_{\text{m}},r_{\text{w}})$ & \textit{Link} & $(\overline{U}_{s,\text{b}},\widetilde{U}_{s,\text{b}})$ & $r_{s,\text{b}}$\tabularnewline
\cmidrule[0.4pt](lr{0.12em}){1-1}%
\cmidrule[0.4pt](lr{0.12em}){2-2}%
\cmidrule[0.4pt](lr{0.12em}){3-3}%
\cmidrule[0.4pt](lr{0.12em}){4-4}%
\cmidrule[0.4pt](lr{0.12em}){5-5}%
\cmidrule[0.4pt](lr{0.12em}){6-6}%
\cmidrule[0.4pt](lr{0.12em}){7-7}%
\cmidrule[0.4pt](lr{0.12em}){8-8}%
vBBU & $\left(2.0,0.20\right)\textrm{e-}4$ & $\left(1.3,0.13\right)\textrm{e-}4$ & $\left(1,0.1\right)\textrm{e-}3$ & $\left(0.4,0.25,2\right)\textrm{e-}2$ & vBBU$\rightarrow$vGW & $\left(1,0.1\right)\textrm{e-}3$ & $0.02$\tabularnewline
vGW & $\left(9.0,0.90\right)\textrm{e-}4$ & $\left(1.3,0.13\right)\textrm{e-}4$ & \textemdash{} & $\left(0.018,0.003,0\right)$ & vGW$\rightarrow$vTM & $\left(1,0.1\right)\textrm{e-}3$ & $0.02$\tabularnewline
vTM & $\left(1.1,0.11\right)\textrm{e-}3$ & $\left(1.3,0.13\right)\textrm{e-}4$ & \textemdash{} & $\left(0.266,0.003,0\right)$ & vTM$\rightarrow$vVOC & $\left(1,0.1\right)\textrm{e-}3$ & $0.02$\tabularnewline
vVOC & $\left(5.4,0.54\right)\textrm{e-}3$ & $\left(3.8,0.38\right)\textrm{e-}3$ & \textemdash{} & $\left(0.108,0.080,0\right)$ & vVOC$\rightarrow$vIDPS & $\left(1,0.1\right)\textrm{e-}3$ & $0.02$\tabularnewline
vIDPS & $\left(1.1,0.11\right)\textrm{e-}2$ & $\left(1.3,0.13\right)\textrm{e-}4$ & \textemdash{} & $\left(0.214,0.003,0\right)$ &  &  & \tabularnewline
\bottomrule
\end{tabular}
\end{table*}
\vspace{-0.8cm}
\par\end{center}

\subsubsection{Background Services\label{dy:subsec:Eva:Background-Services}}

At each infrastructure node $i\in\mathcal{N}$ and link~$ij\in\mathcal{E}$
and for all time slots $k$, we assume that the resources consumed
by best-effort background services follow a normal distribution with
mean and standard deviation equal to respectively $20\%$ and $5\%$
of the available resources at a node and at a link, \textit{i.e}.,
$\left\{ \overline{B}_{n,k}\left(i\right),\widetilde{B}_{n,k}\left(i\right)\right\} =\left\{ 0.2a_{n}\left(i\right),0.05a_{n}\left(i\right)\right\} $
$\forall i\in\mathcal{N}$, $\forall n\in\Upsilon$ and $\left\{ \overline{B}_{\text{b},k}\left(ij\right),\widetilde{B}_{\text{b},k}\left(ij\right)\right\} =\left\{ 0.2a_{\text{b}}\left(ij\right),0.05a_{\text{b}}\left(ij\right)\right\} $,
$\forall ij\in\mathcal{E}$. The reservation impact probability threshold
$\overline{p}^{\text{im}}$ is set to $0.1$.

\subsection{Results\label{dy:subsec:Eva:Results}}

The performance of the reservation variants ($\mathtt{J\text{-}PR}$
and $\mathtt{S\text{-}PR}$) is evaluated considering the following
metrics: slice request acceptance rate, per-slice reservation cost,
average response delay (\textit{i.e}., time between the time instant
the request arrives and the time instant at which it is processed),
average number of adjusted VNF instances per slice, and average computing
time for each slice resource reservation request.

\subsubsection{Resource Reservation for a Single Slice}

The first simulation aims at illustrating the impact of the adaptation
costs described in Section~\ref{subsec:Slice-preparation}, on the
adjustments of the reserved resources between consecutive time slots.
A single slice $s$ is considered. Consequently, the $\mathtt{J\text{-}PR}$
and $\mathtt{S\text{-}PR}$ reservation variants yield the same assignment
$\boldsymbol{\kappa}_{s,\ell}$.\vspace{-0.5cm}

\begin{center}
\begin{figure}[h]
\begin{centering}
\subfloat[$\kappa_{s,\ell}\left(i,v\right)$ when $c_{\text{a}}=0$.\label{dy:fig:Eva:Single:kn:WithoutCa}]{\begin{centering}
\includegraphics[width=0.47\columnwidth]{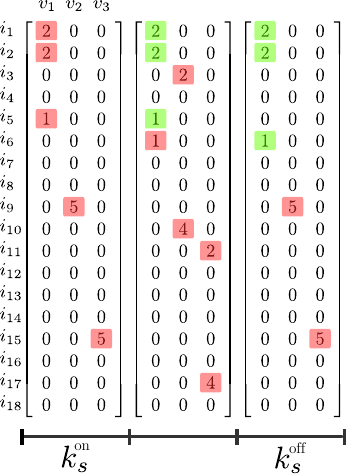}
\par\end{centering}
}\subfloat[$\kappa_{s,\ell}\left(i,v\right)$ when $c_{\text{a}}>0$. \label{dy:fig:Eva:Single:kn:WithCa}]{\begin{centering}
\includegraphics[width=0.47\columnwidth]{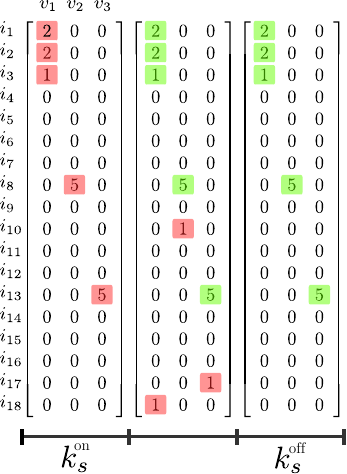}
\par\end{centering}
}
\par\end{centering}
\caption{Evolution of the assignment $\kappa_{s,\ell}\left(i,v\right)$ for
a single slice (for each matrix, rows correspond to $i$, columns
to $v$) when (a) $c_{\text{a}}=0$ and (b) $c_{\text{a}}>0$; the
matrix entries with $\kappa_{s,\ell}\left(i,v\right)-\kappa_{s,\ell\textrm{-}1}\left(i,v\right)>0$
are highlighted in red, whereas entries with $\kappa_{s,\ell}\left(i,v\right)-\kappa_{s,\ell\textrm{-}1}\left(i,v\right)\leqslant0$
and $\kappa_{s,\ell}\left(i,v\right)>0$ are in green. \label{dy:fig:Eva:Single}}
\end{figure}
\vspace{-0.5cm}
\par\end{center}

Figure~\ref{dy:fig:Eva:Single:kn:WithCa} illustrates the evolution
with the time index $\ell$ of $\boldsymbol{\kappa}_{s,\ell}$ for
a slice $s$ of type~1, characterized by an activation duration of
three time slots and a demand pattern of type~2 (increasing for the
second time slot and decreasing for the third one). In Figure~\ref{dy:fig:Eva:Single:kn:WithCa},
the entries for which $y_{s,\ell}\left(i,v\right)=\max\left\{ \kappa_{s,\ell}\left(i,v\right)-\kappa_{s,\ell\textrm{-}1}\left(i,v\right),0\right\} >0$
are highlighted in red, indicating an increase of the reserved resources
for slice~$s$ during consecutive time slots. Comparing Figure~\ref{dy:fig:Eva:Single:kn:WithoutCa},
where $c_{\text{a}}=0$ and Figure~\ref{dy:fig:Eva:Single:kn:WithCa},
where $c_{\text{a}}>0$, one observes that the number of adjustments
of node assignment $\kappa_{s,\ell}\left(i,v\right)$ is reduced when
$c_{\text{a}}>0$, as expected.
\begin{center}
\begin{figure*}[tbh]
\begin{centering}
\subfloat[\label{dy:fig:Eva:Multi:Delay}]{\begin{centering}
\includegraphics[width=0.32\textwidth]{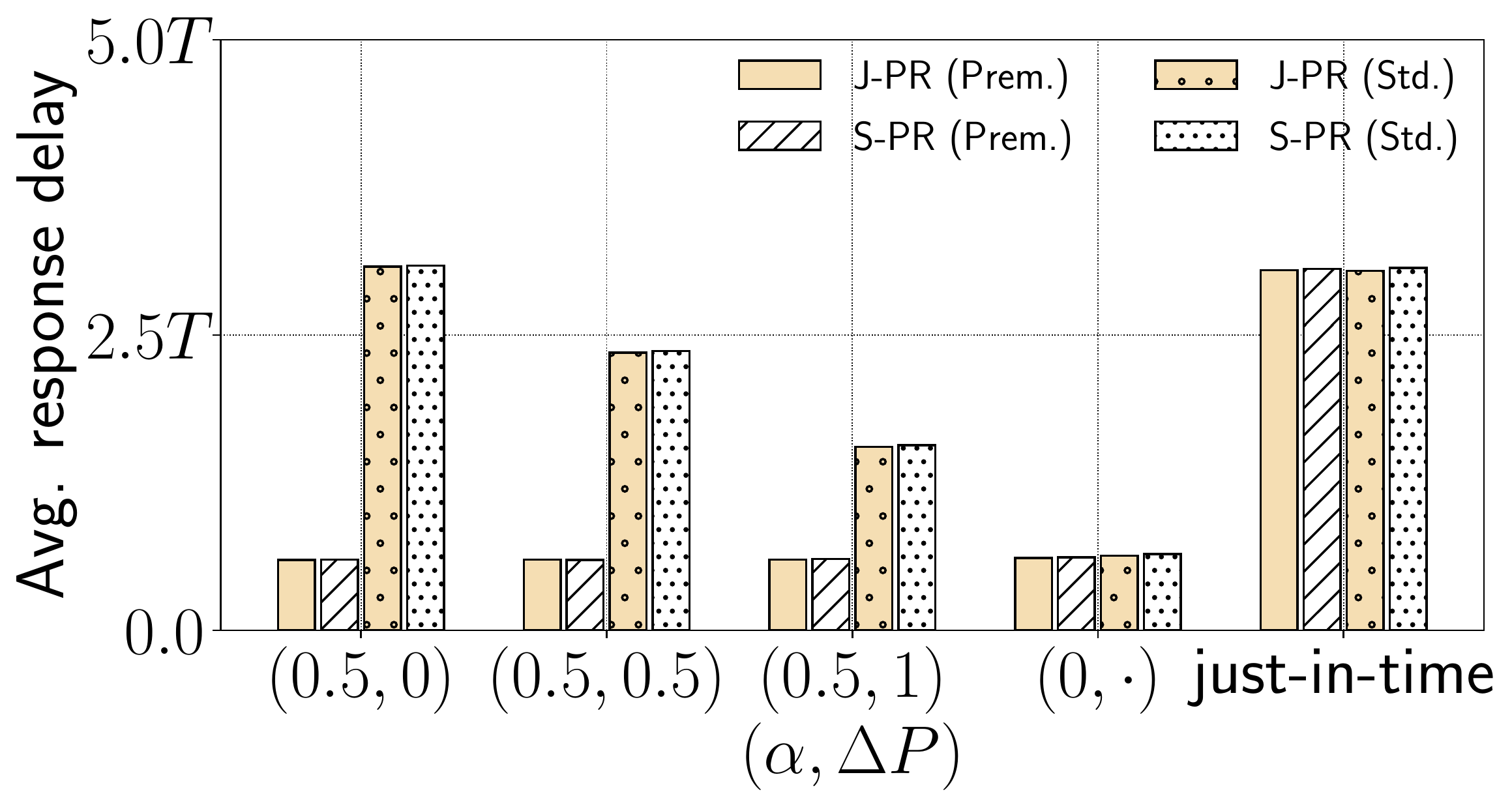}
\par\end{centering}
}\subfloat[\label{dy:fig:Eva:Multi:Rate}]{\begin{centering}
\includegraphics[width=0.32\textwidth]{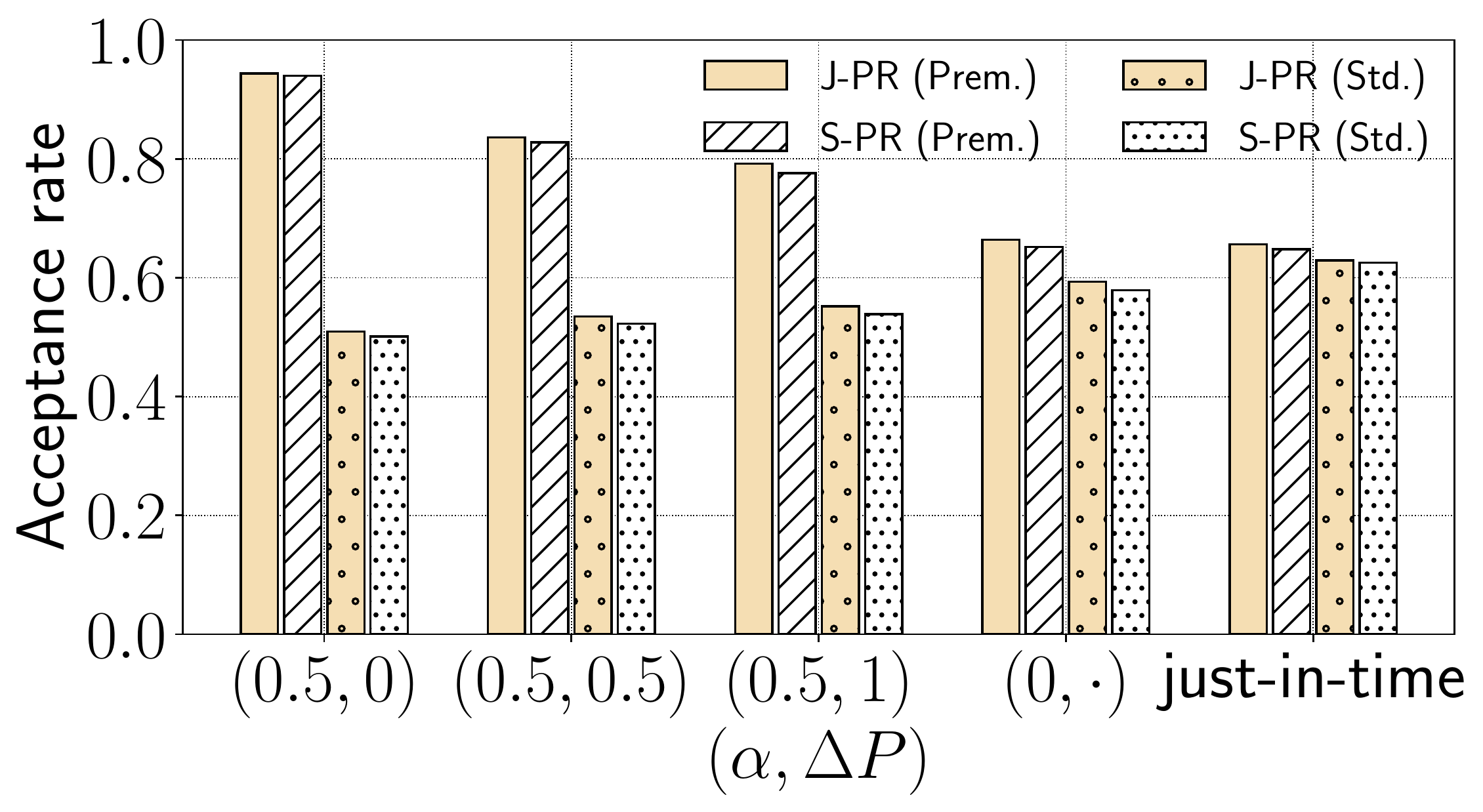}
\par\end{centering}
}\subfloat[\label{dy:fig:Eva:Multi:Moves}]{\begin{centering}
\includegraphics[width=0.32\textwidth]{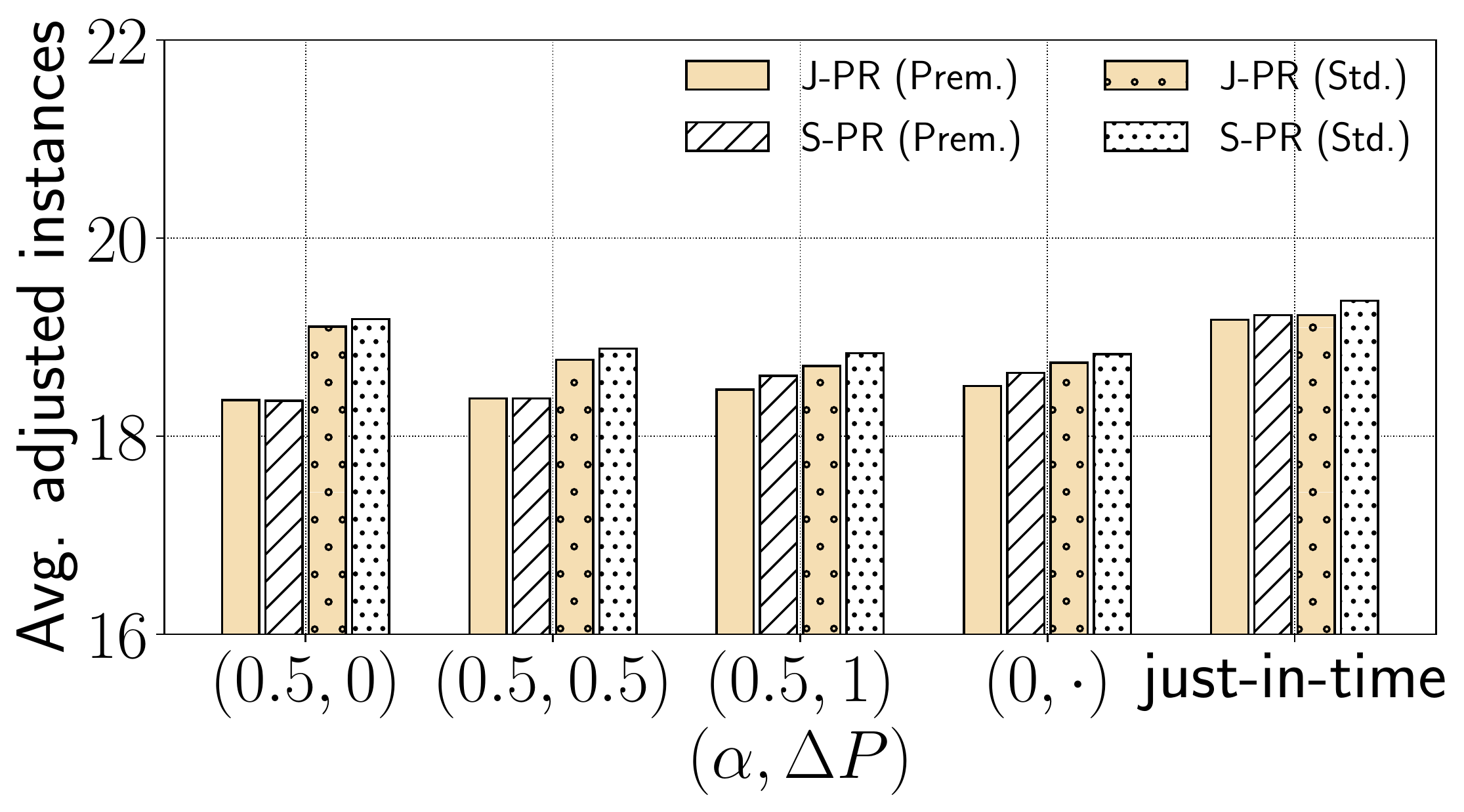}
\par\end{centering}
}\vspace{-0.3cm}
\par\end{centering}
\begin{centering}
\subfloat[\label{dy:fig:Eva:Multi:Cost}]{\begin{centering}
\includegraphics[width=0.32\textwidth]{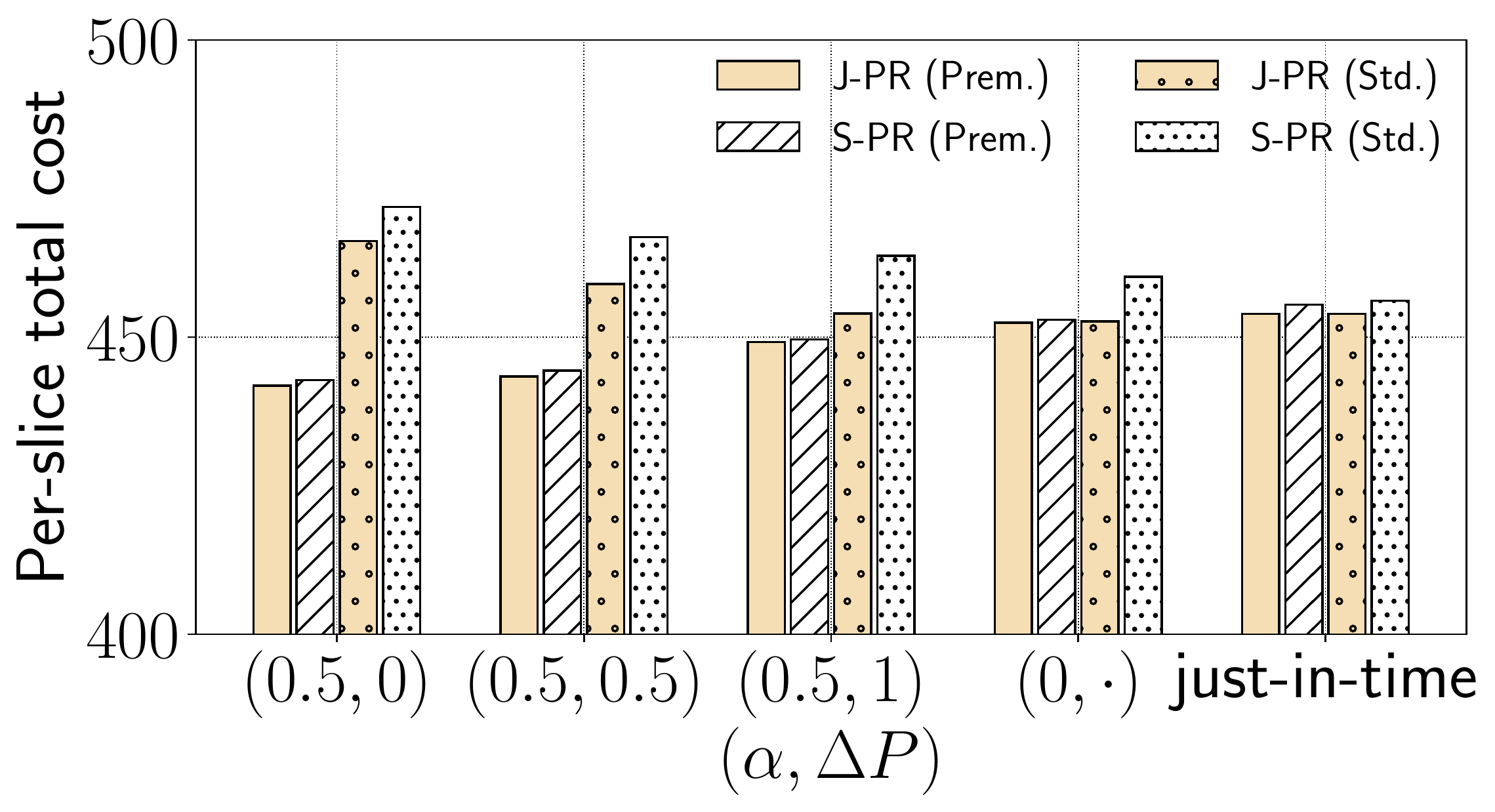}
\par\end{centering}
}\subfloat[\label{dy:fig:Eva:Multi:Time}]{\begin{centering}
\includegraphics[width=0.32\textwidth]{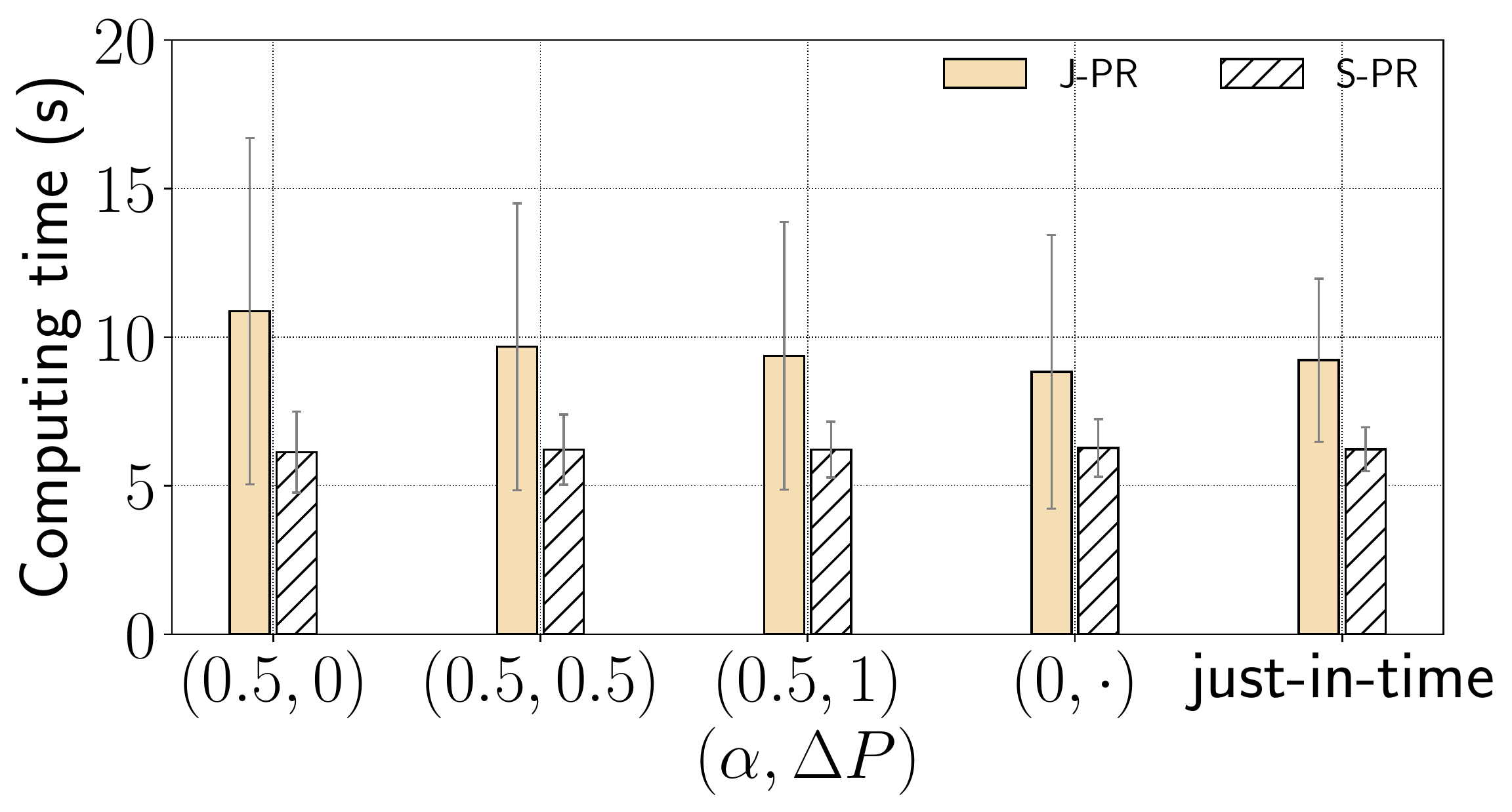}
\par\end{centering}
}
\par\end{centering}
\caption{Performance comparison of the different processing strategies $\left(\alpha,\Delta P\right)$
with the $\mathtt{J\text{-}PR}$ and $\mathtt{S\text{-}PR}$ variants,
in terms of (a) average response delay (expressed as a multiple of
$T$), (b) acceptance rate of slice requests, (b) average adjusted
instances, (d) average cost per slice, and (e) computing time. \label{dy:fig:Eva:Multi}}
\end{figure*}
\vspace{-0.5cm}
\par\end{center}

\subsubsection{Resource Reservation for Multiple Slices}

In this simulation, $1000$ slice requests are generated among which
$250$ are tagged as Premium uniformly at random. Four choices are
considered for the parameters $\alpha$ and $\Delta P$, all with
$P_{\max}=3$, see Section~\ref{dy:subsec:Provisioning-Scheduling}.
These choices impact the processing strategy of Premium and Standard
slice requests. When $\left(\alpha,\Delta P\right)=\left(0.5,0\right)$,
Premium requests are processed immediately and Standard requests are
processed in the time slot preceding their activation time slot. When
$\alpha=0$, whatever the value of $\Delta P$, Premium and Standard
requests are processed immediately, starting with the Premium requests.
With $\left(\alpha,\Delta P\right)=\left(0.5,0.5\right)$ and $\left(\alpha,\Delta P\right)=\left(0.5,1\right)$,
intermediate processing delays are obtained for Standard slices. We
also evaluate the $\mathtt{J\text{-}PR}$ and $\mathtt{S\text{-}PR}$
variants with slice resource reservation requests processed just before
slice activation (\emph{just-in-time} processing). This approach is
close to that considered, \emph{e.g.}, in \cite{Salvat2018}. Nevertheless,
Premium slice requests are still processed first.

Figure~\ref{dy:fig:Eva:Multi} compares the performance of the $\mathtt{J\text{-}PR}$
and $\mathtt{S\text{-}PR}$ reservation variants considering the four
slice requests processing strategies induced by the choices of $\left(\alpha,\Delta P\right)$
and the \emph{just-in-time} approach.

The average response delay (expressed as a multiple of $T$) for each
slice request is shown in Figure~\ref{dy:fig:Eva:Multi:Delay}. The
$\mathtt{J\text{-}PR}$ and $\mathtt{S\text{-}PR}$ variants share
the same prioritized processing policy, therefore, both variants provide
the same result in terms of response delay. When $\alpha=0$, all
requests are processed immediately, independently of their priority.
The observed delay is only due to the processing which takes place
at the end of each time slot during the processing time interval of
duration $\varepsilon T$. When $\alpha=0.5$, the processing delay
remains constant for Premium slices and increases when $\Delta P$
decreases for Standard slices. As expected, the delay is maximum for
the \emph{just-in-time} processing.

Figure~\ref{dy:fig:Eva:Multi:Rate} illustrates the acceptance rate
for the various processing strategies. Processing the slices jointly
yields a slightly higher acceptance rate compared to a sequential
approach. The acceptance rate of Premium slice requests is higher
than that of Standard ones. The difference decreases when the average
processing delay of Standard slice requests decreases. The difference
is minimum when Standard slices are processed just after Premium slices
in the same processing slot, \emph{i.e.}, when $\left(\alpha,\Delta P\right)=\left(0,\cdot\right)$
or with the \emph{just-in-time} approach. Selecting the processing
strategy allows one to adjust the acceptance rate difference between
Premium slices and Standard slices. In practice, the income associated
to both types of slices, as well as the share among these types of
slice requests may help the MNO in determining the best value of the
pair $\left(\alpha,\Delta P\right)$, \emph{e.g.}, that maximizes
its expected income.

Figure~\ref{dy:fig:Eva:Multi:Moves} illustrates the average number
of adjustments of node assignments $y_{s,\ell}\left(i,v\right)$ per
slice and per time slot. A joint approach is again more efficient
than a sequential approach. Moreover, when the processing delay of
Standard slices decreases, the number of adjustments for Standard
slices decreases too, while the average number of adjustments of node
assignments increases for Premium slices. This is explained by the
fact that delaying more the processing of Standard slices facilitates
finding assignments with fewer adjustments during the lifetime of
Premium slices. The price to be paid is more adjustments for Standard
slices. The number of adjustments is maximum with the \textit{just-in-time}
approach.

Figure~\ref{dy:fig:Eva:Multi:Cost} shows the average per-slice reservation
cost charged by the InP to the MNO. Joint resources reservation leads
to lower costs compared to a sequential reservation. The reservation
costs increase for Premium slices when the processing of Standard
slices is less delayed. For Standard and Premium slices, the reservation
costs are consistent with the evolution of the average number of adjustments
of node assignments observed in Figure~\ref{dy:fig:Eva:Multi:Moves}.
The costs for the \emph{just-in-time} approach are similar to those
obtained when the slices requests are processed as they arrive, when
$\left(\alpha,\Delta P\right)=\left(0,\cdot\right)$.

Figure~\ref{dy:fig:Eva:Multi:Time} shows that the computing times
are independent of the processing strategy of Premium and Standard
slice requests. As expected the $\mathtt{S\text{-}PR}$ variant is
less time-consuming than the $\mathtt{J\text{-}PR}$ variant, due
to the reduced number of variables involved. The computing time is
significantly less than the typical time slot duration $T$ when it
is of few tens of minutes. As discussed in Section~\ref{subsec:Complexity-Analysis},
there is a linear relation between the number of nodes and links of
the infrastructure network and the number of variables and constraints
of the max-min optimization problem. The computational complexity
is exponential in the worst case in the number of variables. The fact
that reservation slot duration is of a few tens of minutes allows
InPs to deal with moderate-sized infrastructure networks. Nevertheless,
for large networks, additional heuristics have to be developed to
be able to employ the proposed approach.

\section{Conclusions and Perspectives\label{dy:sec:Conclusions}}

This paper considers a network slicing scenario with slice requests
characterized by variable delays between their submission and activation
and by different priority levels (\emph{e.g.}, Premium and Standard).
Considering these hypotheses, we introduce a prioritized slice resource
reservation and admission control mechanism. Resources required for
admitted slices are reserved, and admission decisions are provided
with a response delay depending on the slice priority and on time
left before its activation.

Adopting the perspective of the InPs, slice resource reservation and
admission control is formulated as a max-min optimization problem.
The InP aims to maximize the number of admitted slices, \emph{i.e.},
slices for which enough resources can be reserved, while minimizing
the cost charged to the MNOs. Uncertainties in the slice resource
demands and the presence of background service sharing the infrastructure
are taken into account. Two reduced-complexity resource reservation
variants, namely $\mathtt{J\text{-}PR}$ and $\mathtt{S\text{-}PR}$,
are proposed to solve the max-min problem.

Numerical results show that the proportion of admitted slices can
be adjusted depending on the difference in the processing delay between
Premium and Standard slices. When the delay difference increases,
resource reservation requests for Premium slices are granted significantly
more frequently, with fewer adjustments with time in the reservation
scheme. This directly impacts the reservation costs, which are reduced
for Premium slices compared to Standard slices when the delay difference
is large. We also illustrate the benefits in terms of satisfaction
differentiation for Premium slice reservation requests provided by
an anticipated processing, compared to a \emph{just-in-time} processing.

The approach presented in this paper may be incorporated in more realistic
simulation environments such as that proposed in \cite{Salvat2018}
to confirm the additional flexibility provided to MNO by differentiated
processing of resource reservation requests.

In the simulations presented in this work, only resource reservation
requests for finite-duration slices have been considered. Periodic
resource reservation requests or requests for slices of very long
duration may be considered without changing the approach. Nevertheless,
this would significantly increase the number of variables to store
the reservation decisions for such requests and would require additional
developments to the proposed approach.

Another possible extension of this work is to enable an adjustment
in future time slots of the resource reservation scheme for slices
that have been admitted, as long as they are not yet activated. In
a given time slot, when Premium slices are processed, often several
assignments lead to the same costs. As seen in \cite{Luu:myo}, accounting
for known slice requests that will have to be processed in future
time slots, may help in the selection among assignments of Premium
slices requests with the same costs, to finally reduce the adaptation
cost of future reservation assignments. Radio coverage constraints
may also be considered in the resource reservation process, using
an approach inspired, \textit{e.g}., from \cite{Luu:cov}. Accounting
for user mobility would also require a model of the mobility patterns
of typical slice users.

\appendices{}

\section{Derivation of Mean and Variance of S-RD \label{dy:sec:Appendix_MeanVariance}}

Consider an active slice~$s$ within time slot $\ell\in\mathcal{K}_{s}$.
The number of users $N_{s,\ell}$ of slice~$s$ and the resource
demands $U_{s,n,\ell}\left(v\right)$ and $U_{s,\text{b},\ell}\left(vw\right)$,
$v\in\mathcal{N}_{s}$, $vw\in\mathcal{E}_{s}$, $n\in\Upsilon$,
of each user of this slice are assumed as independently distributed.
Denoting $\mathbb{E}\left(N_{s,\ell}\right)=\overline{N}_{s,\ell}$,
$\textrm{Var}\left(N_{s,\ell}\right)=\widetilde{N}_{s,\ell}^{2}$,
$\mathbb{E}\left[U_{s,n,\ell}\left(v\right)\right]=\overline{U}_{s,n,\ell}\left(v\right)$,
and $\textrm{Var}\left(U_{s,n,\ell}\left(v\right)\right)=\widetilde{U}_{s,n,\ell}^{2}\left(v\right)$,
the mean value and variance of $R_{s,n,\ell}\left(v\right)$, can
be evaluated, $\forall n\in\Upsilon$ and $\forall v\in\mathcal{N}_{s}$,
as 
\begin{align}
\overline{R}_{s,n,\ell}\left(v\right)= & \mathbb{E}\left(N_{s,\ell}U_{s,n,\ell}\left(v\right)\right)=\overline{N}_{s,\ell}\overline{U}_{s,n,\ell}\left(v\right),\\
\widetilde{R}_{s,n,\ell}^{2}\left(v\right)= & \overline{N}_{s,\ell}^{2}\widetilde{U}_{s,n,\ell}^{2}\left(v\right)+\overline{U}_{s,n,\ell}^{2}\left(v\right)\widetilde{N}_{s,\ell}^{2}\nonumber \\
 & +\widetilde{N}_{s,\ell}^{2}\widetilde{U}_{s,n,\ell}^{2}\left(v\right),
\end{align}
see \cite{Goodman1960}. Similarly, for all $vw\in\mathcal{E}_{s}$,
one obtains 
\begin{align}
\overline{R}_{s,\text{b},\ell}\left(vw\right) & =\overline{N}_{s,\ell}\overline{U}_{s,\text{b},\ell}\left(vw\right),\\
\widetilde{R}_{s,\text{b},\ell}^{2}\left(vw\right) & =\overline{N}_{s,\ell}^{2}\widetilde{U}_{s,\text{b},\ell}^{2}\left(vw\right)+\overline{U}_{s,\text{b},\ell}^{2}\left(vw\right)\widetilde{N}_{s,\ell}^{2}\nonumber \\
 & \quad+\widetilde{N}_{s,\ell}^{2}\widetilde{U}_{s,\text{b},\ell}^{2}\left(vw\right).
\end{align}

\section{Relaxation of the SSP Constraint \label{dy:sec:Appendix_PSP}}

For a given slice $s\in\mathcal{R}_{k}$, , the MNO has to determine
for each time slot $\ell\in\mathcal{K}_{s}$ the smallest value of
$\gamma_{s,\ell}$ such that the satisfaction of \eqref{dy:eq:Relax:Satisfy:Node},
\eqref{dy:eq:Relax:Satisfy:Link} implies that of \eqref{dy:eq:Cplx:Proba:Success}.

Consider the following probability
\begin{equation}
\hspace{-0.5cm}\begin{array}{clcl}
p_{s,\ell}\left(\gamma_{s,\ell}\right)=\Pr\Big\{ & \hspace{-0.35cm}\widehat{R}_{s,n,\ell}\left(v,\gamma_{s,\ell}\right) & \hspace{-0.25cm}\geqslant & \hspace{-0.25cm}R_{s,n,\ell}\left(v\right),\forall v,n,\\
 & \hspace{-0.35cm}\widehat{R}_{s,\text{b},\ell}\left(vw,\gamma_{s,\ell}\right) & \hspace{-0.25cm}\geqslant & \hspace{-0.25cm}R_{s,\text{b},\ell}\left(vw\right),\forall vw\Big\},
\end{array}\label{dy:eq:Pgamma}
\end{equation}
which only involves the slice resource demands as well as $\widehat{R}_{s,n,\ell}\left(v,\gamma_{s,\ell}\right)$
and $\widehat{R}_{s,\text{b},\ell}\left(vw,\gamma_{s,\ell}\right)$
introduced in \eqref{dy:eq:MILP:Bound:Rn} and \eqref{dy:eq:MILP:Bound:Rb}.
For a given value of $\gamma_{s,\ell}$, if $\boldsymbol{\kappa}_{s,\ell}$
is such that \eqref{dy:eq:Relax:Satisfy:Node}, \eqref{dy:eq:Relax:Satisfy:Link}
are satisfied, then, from \eqref{dy:eq:Pgamma}, one has 
\begin{equation}
p_{s,\ell}\left(\boldsymbol{\kappa}_{s,\ell},d_{s}\right)\geqslant p_{s,\ell}\left(\gamma_{s,\ell}\right).
\end{equation}
Consequently, choosing $\gamma_{s,\ell}$ such that $p_{s,\ell}\left(\gamma_{s,\ell}\right)\geqslant\underline{p}_{s}^{\text{sp}}$
implies the satisfaction of \eqref{dy:eq:Cplx:Proba:Success}.

Using \eqref{dy:eq:UserDist} and \eqref{dy:eq:PDF_Rs}, one has
\begin{align}
p_{s,\ell}\left(\gamma_{s,\ell}\right)=\sum\limits _{\eta=1}^{m_{s,\ell}}p_{\eta}{\displaystyle \int_{\mathcal{\widehat{R}}\left(\gamma_{s,\ell}\right)}}f\left(\mathbf{x},\eta\boldsymbol{\mu},\eta^{2}\boldsymbol{\Gamma}\right)\textrm{d}\mathbf{x},\label{dy:eq:Relax_Proba_Success}
\end{align}
where $\mathcal{\widehat{R}}\left(\gamma_{s,\ell}\right)=\left\{ \mathbf{x}\in\mathbb{R}^{n_{s}}\,|\,\mathbf{x}\leqslant\widehat{\mathbf{R}}\left(\gamma_{s,\ell}\right)\right\} $
with 
\begin{align}
\widehat{\mathbf{R}}\left(\gamma_{s,\ell}\right)=\big( & \widehat{R}_{s,\text{c},\ell}\left(v_{1},\gamma_{s,\ell}\right),\widehat{R}_{s,\text{m},\ell}\left(v_{1},\gamma_{s,\ell}\right),\dots\nonumber \\
 & \dots,\widehat{R}_{s,\text{b},\ell}\left(v_{1}v_{2},\gamma_{s,\ell}\right),\dots\big)^{\top}
\end{align}
of size $n_{s}=\left|\Upsilon\right|\left|\mathcal{N}_{s}\right|+\left|\mathcal{E}_{s}\right|$.
If the pmf $p_{\eta}$ of the number $N_{s,\ell}$ of users is known,
the value of $\gamma_{s,\ell}$ such that $p_{s,\ell}\left(\gamma_{s,\ell}\right)=\underline{p}_{s}^{\text{sp}}$
can be obtained by bisection search methods, see, \emph{e.g.}, \cite{Burden2011}.
The multidimensional integral in \eqref{dy:eq:Relax_Proba_Success}
can be evaluated, \emph{e.g.}, using a quasi-Monte Carlo integration
algorithm presented in \cite{Genz2004}.

The evaluation of $\gamma_{s,k}$ can be performed as soon as the
reservation request for slice $s$ is received, prior to the optimization
of the reservation.

\section{Relaxation of the IP Constraint \label{dy:sec:Appendix_IP}}

For a given slice $s\in\mathcal{R}_{k}$ and an assignment $\boldsymbol{\kappa}_{s,\ell}$,
$\ell\in\mathcal{K}_{s}$ that satisfies (\ref{dy:eq:Relax:Limit:Node},
\ref{dy:eq:Relax:Limit:Link}) and (\ref{dy:eq:Cplx:Cons:Limit:Node},
\ref{dy:eq:Cplx:Cons:Limit:Link}, \ref{dy:eq:Cplx:Cons:Flow}), the
IP defined in \eqref{dy:eq:Cplx:Proba:Impact:Node} can be evaluated
as, $\forall i\in\mathcal{N}$, $\forall n\in\Upsilon$,
\begin{align}
p_{n,\ell}^{\text{im}}\left(i\right) & =\Pr\Big\{ B_{n,\ell}\left(i\right)\geqslant\widehat{B}_{n,\ell}\left(i,\gamma_{\text{B},\ell}\right)\Big\}\nonumber \\
 & =1-\int_{-\infty}^{\widehat{B}_{n,\ell}\left(i,\gamma_{\text{B},\ell}\right)}\hspace{-0.2cm}f\left(x;\overline{B}_{n,\ell}\left(i\right),\widetilde{B}_{n,\ell}^{2}\left(i\right)\right)\textrm{d}x\nonumber \\
 & =1-\Phi\left(\gamma_{\text{B},\ell}\right),\label{dy:eq:Relax_Proba_Impact_Node}
\end{align}
where $\Phi$ is the cumulative distribution function (CDF) of the
zero-mean, unit-variance normal distribution. Similarly, the IP defined
in \eqref{dy:eq:Cplx:Proba:Impact:Link} can also be evaluated $\forall ij\in\mathcal{E}$
as, 
\begin{align}
\hspace{-0.5cm}p_{s,\text{b},\ell}^{\text{im}}\left(ij\right) & =\Pr\Big\{ B_{\text{b},\ell}\left(ij\right)\geqslant\widehat{B}_{\text{b},\ell}\left(ij,\gamma_{\text{B},\ell}\right)\Big\}\nonumber \\
 & =1-\Phi\left(\gamma_{\text{B},\ell}\right).\label{dy:eq:Relax_Proba_Impact_Link}
\end{align}
Both \eqref{dy:eq:Relax_Proba_Impact_Node} and \eqref{dy:eq:Relax_Proba_Impact_Link}
are independent of $\boldsymbol{\kappa}_{s,\ell}$, $\forall s\in\mathcal{R}_{k}$.
To satisfy the impact constraints imposed by (\ref{dy:eq:Cplx:Proba:Impact:Node},
\ref{dy:eq:Cplx:Proba:Impact:Link}), $\gamma_{\text{B},\ell}$ has
to be chosen such that $1-\Phi\left(\gamma_{\text{B},\ell}\right)\leqslant\overline{p}^{\text{im}}$,
\emph{i.e.}, such that
\begin{align}
\gamma_{\text{B},\ell}\geqslant\Phi^{-1}\left(1-\overline{p}^{\text{im}}\right).\label{dy:eq:Relax_Gamma_B}
\end{align}
Since the larger $\gamma_{\text{B},\ell}$, the more difficult the
satisfaction of \eqref{dy:eq:Relax:Limit:Node} and \eqref{dy:eq:Relax:Limit:Link},
the optimal $\gamma_{\text{B},\ell}$ would be $\gamma_{\text{B},\ell}=\Phi^{-1}\left(1-\overline{p}^{\text{im}}\right)$.

\vskip -1\baselineskip plus -1fil
{\scriptsize
\bibliographystyle{IEEEtran} 
\bibliography{ref_dyn}
}

\vskip -2\baselineskip plus -1fil
\begin{IEEEbiography}[{\includegraphics[width=1in,height=1.25in]{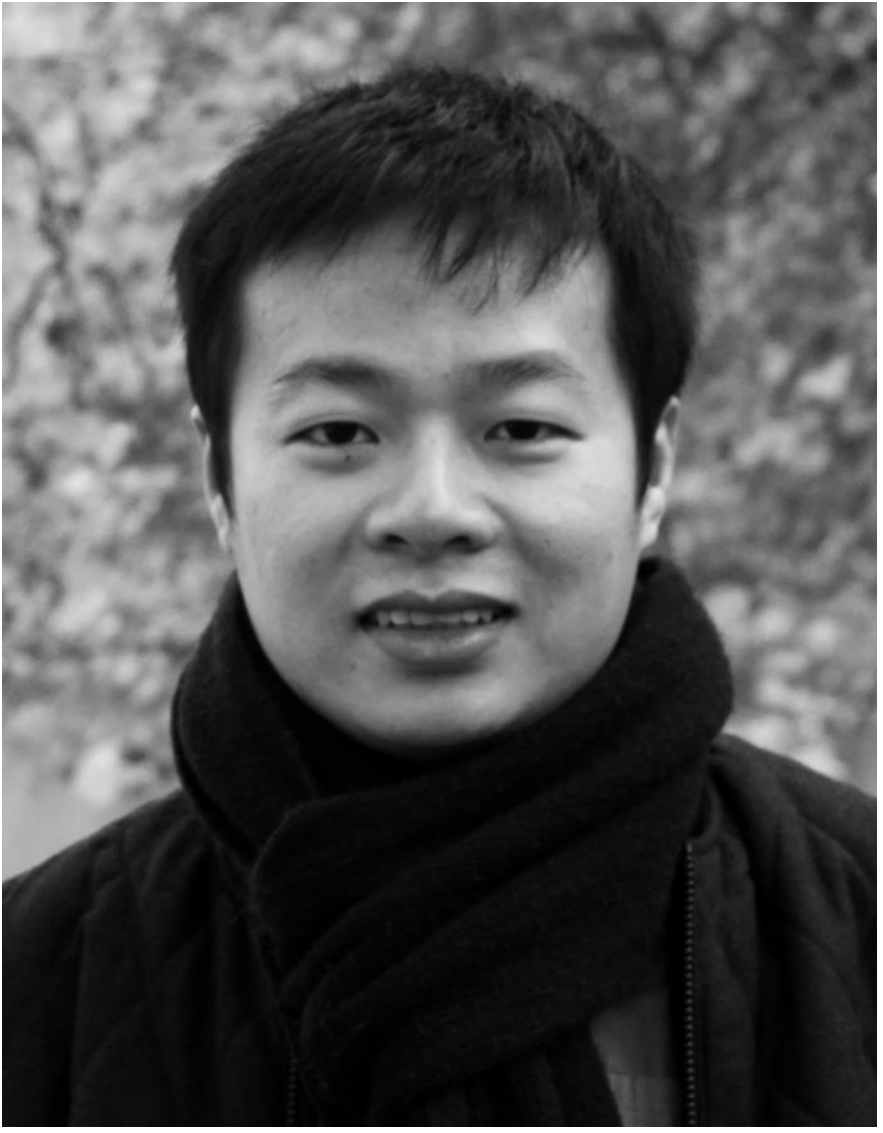}}]{Quang-Trung Luu}is currently a postdoctoral fellow at the Laboratory for Analysis
and Architecture of Systems (LAAS), French National Centre for Scientific
Research (CNRS), Toulouse, France. He received a Ph.D in networking
and telecommunications from Paris-Saclay University, France in 2021.
During the Ph.D, he was also a research engineer at Nokia Bell Labs,
France. His research focuses on the optimization of resource management
in communication networks, in particular on key enabling technologies
for next-generation mobile systems.

\end{IEEEbiography}
\vskip -2.8\baselineskip plus -1fil
\begin{IEEEbiography}[{\includegraphics[width=1in,height=1.25in]{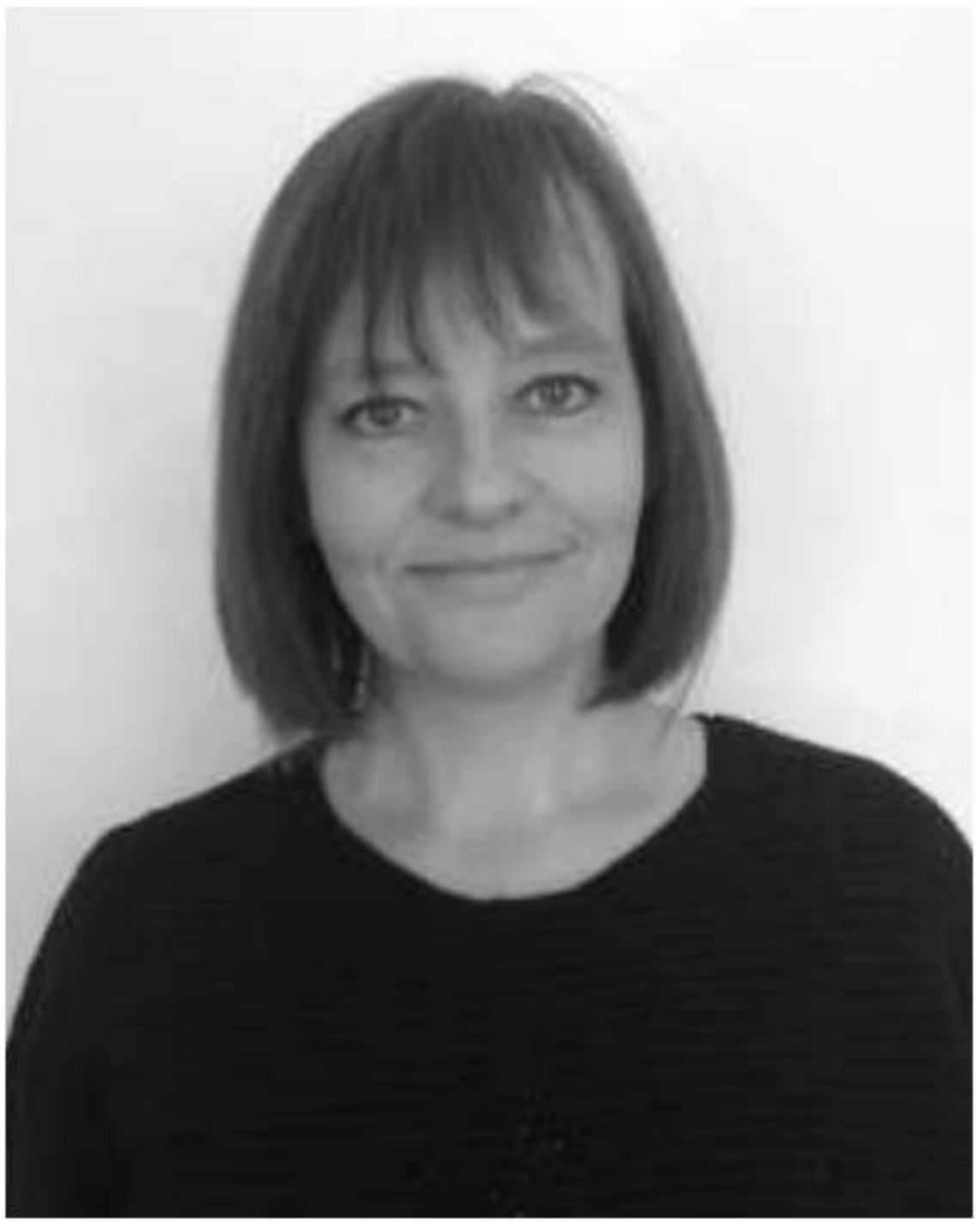}}]{Sylvaine Kerboeuf} received the M.S. degree in physics and the Ph.D. degree in solid-state
physics from the University of Paris-Sud, Orsay, France, in 1991 and
1994, respectively, and the Ph.D. degree in superconductivity from
the Centre National d'Etude des T\'el\'ecommunications, France Telecom,
Paris, France. She joined the Research and Innovation Department,
Alcatel-Lucent Bell Labs, Nozay, France, where she was involved in
research projects on optoelectronics for several years. In 2004, she
joined a project involved in radio access networks and focusing on
fourth generation discontinuous networks and on caching technology.
She is currently a Senior Researcher in the Wireless Program with
Nokia Bell Labs. Her current research interests include software defined
network architecture, network slicing and end-to-end orchestration
of micro-services for 5G networks.

\end{IEEEbiography}
\vskip -2.8\baselineskip plus -1fil
\begin{IEEEbiography}[{\includegraphics[width=1in,height=1.25in]{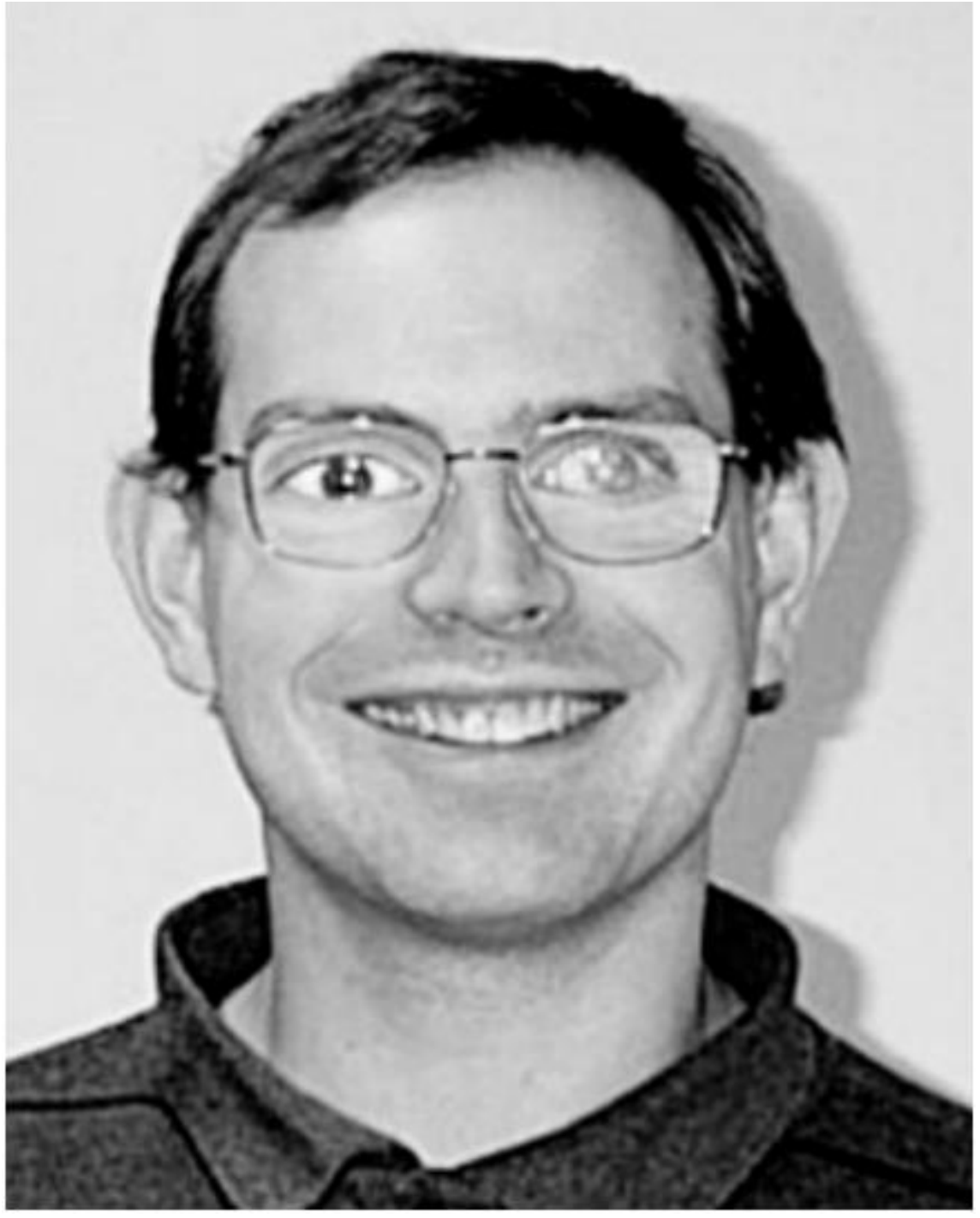}}]{Michel Kieffer} (M'02, SM'07) received the Ph.D. degree in control theory from the
University of Paris XI, Orsay, France, in 1999. He is a Full Professor
in signal processing for communications with the University of Paris-Sud
and a Researcher with the Laboratoire des Signaux et Syst{\`e}mes
(L2S), Gif-sur-Yvette, France. Since 2009, he is a part-time Invited
Professor with the Laboratoire Traitement et Communication de l'Information,
T\'el\'ecom ParisTech, Paris, France. He is coauthor of more than
150 contributions in journals, conference proceedings, or books. He
is one of the coauthors of the books \textit{Applied Interval Analysis}
(Springer-Verlag, 2001) (this book was translated in Russian in 2005)
and \textit{Joint Source-Channel Decoding: A Cross-Layer Perspective
With Applications in Video Broadcasting} (Academic, 2009). His research
interests are in signal processing for multimedia, communications,
and networking; distributed source coding; network coding; joint source-channel
coding and decoding techniques; and joint source-network coding. Applications
are mainly in the reliable delivery of multimedia contents over wireless
channels. He is also interested in guaranteed and robust parameter
and state bounding for systems described by nonlinear models in a
bounded-error context. Prof.~Kieffer was a junior member of the \textit{Institut
Universitaire de France} from 2011 to 2016. He serves as an Associate
Editor of $\textsc{Signal Processing}$ since 2008 and of the $\textsc{IEEE Transactions on Communications}$
from 2012 to 2016.

\end{IEEEbiography}
\end{document}